\documentclass[%
superscriptaddress,
preprint,
amsmath,amssymb,
aps,
prl,
]{revtex4-2}
\usepackage{graphicx}
\usepackage{multirow}
\usepackage{color}
\usepackage{times}
\usepackage{amsmath,bm,amsfonts}
\usepackage{url}
\usepackage{tikz}
\usepackage{dsfont}
\usepackage{mathtools}
\usepackage{dcolumn}
\usepackage{setspace}
\usepackage{ulem}
\usepackage{physics}
\usepackage{academicons}

\usepackage[toc,page,titletoc]{appendix}

\usepackage[colorlinks=true,linkcolor=black,citecolor=black]{hyperref}
\usepackage{orcidlink}
\hypersetup{
    linkcolor=black,       % Color of internal links
    citecolor=black,       % Color of citations
    filecolor=black,       % Color of file links
    urlcolor=black         % Color of URLs
}
\usepackage{orcidlink}

\def\ba#1\ea{\begin{align}#1\end{align}}
\def\bg#1\eg{\begin{gather}#1\end{gather}}
\def\bpm{\begin{pmatrix}}
	\def\epm{\end{pmatrix}}

\newcommand{\blue}[1]{\textcolor{black}{#1}}

\begin{document}
	%	%%%%%%%%%%%%%TITLE%%%%%%%%%%%%%
	\title{Quasiperiodic pairing in graphene quasicrystals}
	%%%%%%%%%%%%%%%%%%%%%%%%%%%%%%%
	
	%%%%%%%%%%%%AUTHORS%%%%%%%%%%%%
	\author{Rasoul Ghadimi}
	% \orcidlink{0000-0001-6927-4008}
	\affiliation{Department of Physics and Astronomy, Seoul National University, Seoul 08826, Korea}
	\affiliation{Center for Theoretical Physics (CTP), Seoul National University, Seoul 08826, Korea}
	\affiliation{ Institute of Applied Physics, Seoul National University, Seoul 08826, Korea}

	\author{Bohm-Jung Yang}
	\email{bjyang@snu.ac.kr}
	\affiliation{Department of Physics and Astronomy, Seoul National University, Seoul 08826, Korea}
	\affiliation{Center for Theoretical Physics (CTP), Seoul National University, Seoul 08826, Korea}
	\affiliation{ Institute of Applied Physics, Seoul National University, Seoul 08826, Korea}
	%%%%%%%%%%%%%%%%%%%%%%%%%%%%%%%
	\date{\today}
	%%%%%%%%%%%ABSTRACT%%%%%%%%%%%%
	\begin{abstract}
We investigate the superconducting instabilities of twisted bilayer graphene quasicrystals (TBGQC) obtained by stacking two monolayer graphene sheets with a $30^\circ$ relative twisting.
The electronic energy spectrum of TBGQC contains periodic energy ranges (PER) and quasiperiodic energy ranges (QER), where the underlying local density of states (LDOS) exhibits periodic and quasiperiodic distribution, respectively.
We found that superconductivity in the PER is a simple superposition of two monolayer superconductors. This is because, particularly near the charge neutrality point of TBGQC, the two layers are weekly coupled, leading to pairing instabilities with uniform distribution in real space.
On the other hand, within QER, the inhomogeneous distribution of the LDOS enhances the superconducting instability with a non-uniform distribution of pairing amplitudes, leading to quasiperiodic superconductivity. Our study can qualitatively explain the superconductivity in recently discovered moir\'e quasicrystals, which show superconductivity in its QER.
	\end{abstract}

	\date{\today}
	\maketitle
	
	%%%%%%%%%%%%%BACKGROUND%%%%%%%%%%%%%%%%%%

	Quasicrystals \cite{PhysRevLett.53.2477} (QCs) stand as an intermediary between disordered and periodic structures, giving rise to a unique blend of physics derived from both realms   \cite{PhysRevLett.53.1951,PhysRevB.34.596,senechal1996quasicrystals}. 
	Consequently, they exhibit captivating properties such as quantum criticality   \cite{Deguchi2012}, confined states   \cite{PhysRevB.37.2797,PhysRevB.38.1621,PhysRevB.96.214402,PhysRevB.102.064213,PhysRevB.105.045146,PhysRevB.108.125104}, self-similarity/fractals   \cite{PhysRevB.35.1020,doi:10.7566/JPSJ.86.114707,PhysRevLett.122.110404,PhysRevB.108.064210}, and higher-dimensionality   \cite{PhysRevLett.111.226401,PhysRevB.106.L201113,PhysRevResearch.4.013028}. 
	The absence of translational symmetry in QC, despite their long-range patterns, opens up many questions regarding the emergence of conventional crystalline phenomena, such as topology   \cite{PhysRevLett.116.257002,PhysRevLett.123.196401,Layer_dependent_topological_quasicrystals_Jeffrey2020,PhysRevResearch.2.033071,RevModPhys.93.045001,Topological_quasicrystals_Fan2021,Lv2021_quasicrystalline_quadrupole,PhysRevResearch.6.033088,PhysRevB.109.174512,PhysRevB.102.241102,PhysRevLett.124.036803},  correlated states   \cite{PhysRevA.72.053607,PhysRevA.100.053609,PhysRevB.102.224201,PhysRevLett.126.110401,PhysRevResearch.3.023180,PhysRevLett.131.173402}, and magnetism   \cite{PhysRevLett.90.177205,PhysRevB.75.212407,PhysRevB.77.104427,Deguchi2012,PhysRevLett.115.036403,PhysRevMaterials.4.024417,PhysRevMaterials.4.024417,Watanabe2021}. 
	Furthermore, despite their quasiperiodicity, QCs exhibit a trace of energy dispersion in their spectral function, suggesting the presence of partially Bloch-like wavefunctions   \cite{PhysRevB.100.081405,PhysRevA.97.043603,PhysRevB.104.144511,PhysRevResearch.4.013226,doi:10.7566/JPSCP.38.011062}, which gives rise to exotic phenomena including quantum oscillations \cite{PhysRevB.100.081405} as in periodic metallic systems.

	Interestingly, in line with the existence of many intriguing crystalline phenomena in QCs, superconductivity has recently been discovered also in QC materials \cite{Kamiya2018,Tokumoto2024}.
	Theoretically, superconductivity in QC can arise either intrinsically  \cite{PhysRevB.95.024509,PhysRevB.100.014510,PhysRevLett.125.017002,  doi:10.7566/JPSCP.38.011065} or through the proximity effect  \cite{PhysRevB.100.165121,PhysRevB.102.134211}.   
	Despite the absence of well-defined Bloch wavefunctions, modified versions of crystalline superconductivity theories \cite{ANDERSON195926} were proposed to explain superconductivity in QCs \cite{PhysRevB.100.014510}.
	However, the quasiperiodicity effect can strongly affect superconducting properties in QC  \cite{  PhysRevB.102.115108, PhysRevB.104.144511,PhysRevResearch.3.023195,PhysRevResearch.3.013262,PhysRevB.106.064506,PhysRevResearch.5.043164, Liu2023, PhysRevB.110.144512,PhysRevB.110.L060501,PhysRevB.108.054203,PhysRevB.109.134504}. 
	For instance, superconductivity in QC can potentially give a ``fractal superconductor" due to the underlying self-similar patterns  \cite{Kamiya2018}.
	Interestingly, despite the strong dependence of the order parameter on local environments  \cite{PhysRevB.102.224201}, the superconducting gap remains uniform throughout the system with a well-defined critical temperature \cite{PhysRevB.109.214507}. 
	However, specific heat jumps at the superconducting transition and Bogoliubov peaks may be suppressed due to the underlying quasiperiodic wavefunction \cite{PhysRevB.102.115108}. 
    Furthermore, the cooperative effect of correlation length divergence at the phase transition and the scaling behavior of the local environment in QC introduces a new scaling law for the aperiodical order parameter \cite{PhysRevB.102.224201}. Additionally, QC superconductivity exhibits distinct supercurrent responses  \cite{Liu2022,PhysRevResearch.5.043164,PhysRevLett.125.017002}.

	Recent advances in the study of twisted multilayer moiré materials have garnered new attention to extrinsic QCs \cite{PhysRevB.99.165430}.
	Unlike intrinsic QCs with inherent quasiperiodicity, extrinsic QCs originate from stacking two-dimensional periodic lattices with relatively incommensurate basis vectors.
	In certain cases, this stacking leads to atomic QC structures with a quasiperiodic tiling of clusters of atoms. 
	In particular, twisting two graphene layers with a $30^\circ$ angle leads to twisted bilayer graphene quasicrystal (TBGQC), which manifests 12-fold dodecagonal Stampfli tiling   \cite{stampfli1986dodecagonal,PhysRevB.93.201404}. 
	The TBGQC can commonly appear alongside more energetically favorable AB stacked bilayer graphene  \cite{doi:10.1021/acs.chemmater.6b01137}.
	Furthermore,  TBGQC was experimentally confirmed by observing low-energy Dirac cones and their mirrored images due to Umklapp interlayer scattering. 
 It was suggested that these Dirac cones may give rise to relativistic Dirac physics within QC realms   \cite{doi:10.1126/science.aar8412,doi:10.1073/pnas.1720865115,Yan_2019}. 
	However, it was later realized that these Dirac cone replicas do not contribute to the electrical transport   \cite{doi:10.1021/acs.nanolett.0c00172}, and effectively low energy electronic properties are described by the original graphene's Dirac cones   \cite{doi:10.1021/acsnano.9b07091}. 
 Moreover, huge interlayer coupling is needed to achieve quasicrystal physics at the Fermi energy   \cite{PhysRevB.99.245401,PhysRevB.104.165112}.
To avoid such unrealistic conditions, the Fermi energy must be shifted away from the Dirac point to the higher energy window, where new van Hove singularities (vHS) with quasicrystalline wavefunctions appear \cite{PhysRevB.99.165430,PhysRevB.103.045408}.
	Therefore, TBGQC hosts both periodic energy ranges (PER) and quasiperiodic energy ranges (QER) simultaneously, and quasiperiodicity can be toggled on or off by external stimuli such as pressure, and electrostatic gating
	\cite{PhysRevB.102.045113}.

	In this Letter, motivated by the recent discoveries of superconductivity in extrinsic QCs~\cite{Uri2023}, we study the superconducting instability of TBGQC, in a broad energy regime including both the PER and QER.
	We note that superconductivity has been observed in both twisted or non-twisted multi-layer graphene, particularly when the electronic dispersion is flattened \cite{Cao2018,Park2021,Zhou2021,Oh2021,doi:10.1126/science.abm8386,Pantaleón2023}.
	In the case of TBGQC, the tunability of its quasiperiodicity using Fermi energy manipulation offers a unique opportunity to study the quasiperiodicity effect on QC superconductors.
		It has been shown that TBGQC may host high-angular-momentum topological superconductivity for PER \cite{PhysRevB.107.014501}. However, the significance and effects of quasiperiodicity on superconducting instability in QER still remain elusive. 
	We found that for Fermi energy in  PER, the pairing instability can be understood as a simple superposition of two monolayer graphene superconductors.
	On the other hand, in QER, the quasiperiodicity not only induces the quasiperiodic modulation of pairing amplitudes but also enhances the superconductivity itself, especially for onsite spin-singlet pairing.
	Our work paves a way to understand the fundamental properties of the emergent superconductivity in various systems with inherent quasiperiodicity such as TBGQC and moir\'e quasicrystals~\cite{Uri2023}.

%%%%%%%%%%%%%%%%%%%%%%%%%%%%%%%%%%%%%%%%%%%%%%%%%%%%%%%	
	\begin{figure}[t!]
		\includegraphics[width=1\linewidth]{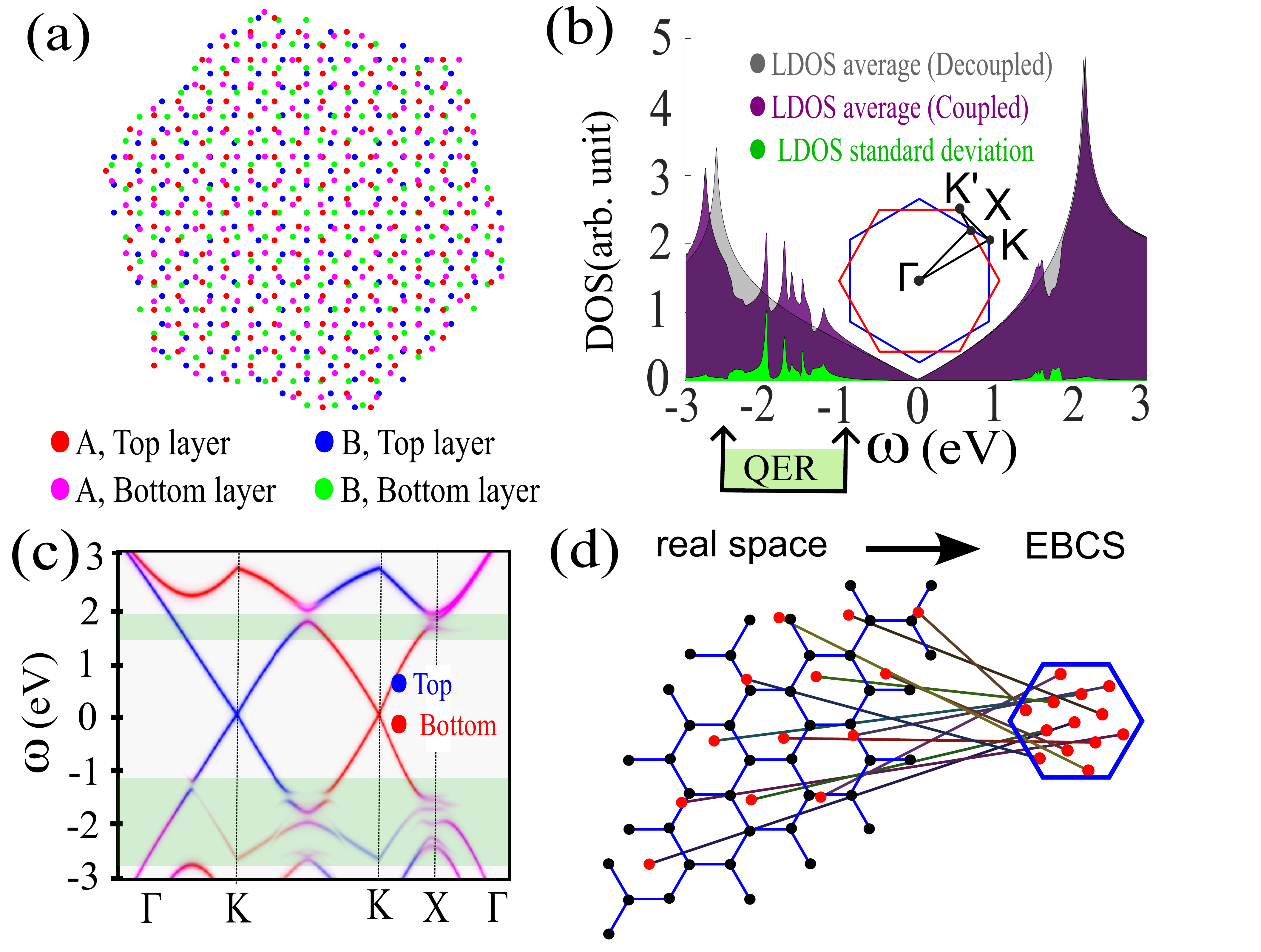}
		\caption{(a) A super-cell of twisted bilayer graphene quasicrystal (TBGQC)  approximant for $\theta\approx 30.1583$ [$\tau_D(4)=7/4$].
			(b) The averaged local density of states (LDOS) $\rho(\omega)$ for the bilayer with (purple) and without (grey) inter-layer coupling.
			The green plot shows the standard deviation $\sigma_{\rho}(\omega)$ of the LDOS.
			The red and blue hexagons indicate the first Brillouin zone of two decoupled graphene layers.
			(c) Unfolded band structure along the path indicated in (b), where the blue and red show the spectral function mainly dominated by the top and bottom layers, respectively. 
			The green-shaded energy range shows the region with considerable LDOS fluctuations.
			(d) Environment-based classification space (EBCS) map for TBGQC. 
        In EBCS, vertices with similar environments are mapped to neighboring positions within a hexagon. 
			% To obtain an EBCS, the relative positions of all the projected lattice points within their enclosing hexagons are collected in a single hexagon where the vertices with similar environments are located at neighboring positions. 
		}
		\label{fig:fig1}
	\end{figure}
%%%%%%%%%%%%%%%%%%%%%%%%%%%%%%%%%%%%%%%%%%%%%%%%%%%%%	
	
	To elucidate the irrational structure of TBGQC, let us consider a general TBG. 
	The basis lattice vectors of the first layer are given by $\vec{a}_{1,2}= a (\tfrac{3}{2},\pm\tfrac{\sqrt{3}}{2})$, and the corresponding second layer vectors $\vec{a}'_{1,2}$ are obtained by $\theta$ rotation, where $a$ is the shortest carbon-carbon distance. 
	The TBG can form a commensurate structure when $\cos\theta=\frac{3p^2+3pq+\frac{q^2}{2}}{3p^2+3pq+q^2}$ holds where the integer $p$ and $q$ satisfy $p \vec{a}_2 + (p + q) \vec{a}_1=(p+q) \vec{a}'_2 + p \vec{a}'_1$ ~\cite{PhysRevB.86.155449}.
	Therefore, the twisted layers make a super-lattice with larger basis vectors  $\vec{A}_1=p \vec{a}_2 + (p + q) \vec{a}_1$, $\vec{A}_2=(2p+q) \vec{a}_2 - (p + q) \vec{a}_1$, and its super-cell containing $N_{\text{cell}}=4(3p^2+3 p q+q^2)$ atoms. 
	The TBGQC is obtained by choosing $\theta=30^\circ$, at which $p$ and $q$ satisfy $\tau_D\equiv\tfrac{q}{p}=\sqrt{3}$.
	The irrational number $\tau_D$ represents the quasi-periodicity of TBGQC similar to the golden ratio in Penrose QC~\cite{PhysRevB.106.L201113,PhysRevB.96.214402}.

	Analyzing QC structures can be simplified by studying their corresponding periodic approximants~\cite{PhysRevB.43.8879}.
	To find different approximants of TBGQC ~\cite{Yu2019}, one can approximate $\tau_D$ by truncating the continued fraction of $\sqrt{3}$ and obtain two coprime integers $p_g$ and $q_g$ satisfying $\tau_D(g)=\tfrac{q_g}{p_g}$ where $g$ is an index for approximant generations, and by definition satisfies $\lim_{g\rightarrow\infty}\tau_D(g)=\sqrt{3}$.
	For instance, selecting values like $\tau_D(1)= \tfrac{1}{1}, \tau_D(2)=\tfrac{2}{1},  \tau_D(3)=\tfrac{5}{3}$, and $\tau_D(4)=\tfrac{7}{4}$ results in approximants containing $N_{\text{cell}} = 28, 52, 388,$ and $724$ atoms, respectively. 
	In this work, our primary focus is  g=4 ($\tau_D(4){=}\tfrac{7}{4}$),  or equivalently $\theta = 30.1583^\circ$.
	Fig.~\ref{fig:fig1}(a) shows the corresponding approximant supercell, which is large enough to capture the quasiperiodicity while ensuring computational feasibility.
	
	To describe the energy spectrum of TBGQC, we use a  tight-binding Hamiltonian that describes hopping between carbon $p_z$ orbitals~\cite{PhysRevB.102.115123} \blue{(see the Supporting Information [S1])}. 
	The translational symmetry of the TBGQC approximant allows us to diagonalize the Hamiltonian in the supercell momentum space, which gives the energy eigenvalues $\epsilon_{\vec k, \lambda}$ ($\lambda=1, \dots, N_{\text{cell}}$) and the energy eigenvectors $\ket{\psi_{\vec k,\lambda}}$.
	In Fig.~\ref{fig:fig1}(b), the average local density of states (LDOS) $\rho(\omega)\equiv\tfrac{1}{N_{\text{cell}}}\sum_{i=1}^{N_{\text{cell}}}\rho_i(\omega)$ is plotted with (purple plot) and without (grey plot) interlayer coupling, respectively. 
	Here, $\rho_i(\omega)=\sum_{\vec k,\lambda} \left|\braket{i}{\psi_{\vec k,\lambda}}\right|^2 \delta(\epsilon_{\vec k, \lambda}-\omega)$ is the LDOS for the $i$-th site in a supercell. 
	We note that the interlayer coupling generates singular peaks in LDOS, which arise from new vHS in TBGQC, dubbed the quasicrystal vHS (QCvHS), in the energy range of $\omega\in[-2.5,-1] \text{eV}$~\cite{PhysRevB.99.165430}.
	To examine the influence of quasiperiodicity on the wavefunction, we show [green plot in Fig.~\ref{fig:fig1}(b)] the standard deviation of the LDOS, $\sigma_{\rho}(\omega)\equiv \frac{1}{N_{\text{cell}}}\sqrt{\sum_i(\rho_i(\omega)-\rho(\omega))^2}$.
	$\sigma_{\rho}(\omega)$ is prominent near the newly generated QCvHS, indicating the importance of the underlying QC structure for these energy ranges. 
	In a previous study, the inverse participation ratio (IPR) enhancement was observed in a similar energy range, indicating the non-uniformity of the corresponding wavefunctions~\cite{PhysRevB.99.165430}. 
	In the subsequent discussion, we refer to the energy range $\omega\in[-2.5,-1] \text{eV}$ with substantial LDOS fluctuations as QER [indicated in Fig.~\ref{fig:fig1}(b)].
	
	Since the TBGQC approximant contains many atoms, the resulting band structure is complicated and contains many folded energy bands of original graphene. 
	However, one can unfold the band structure by calculating spectral functions \cite{PhysRevB.102.115123}, which captures how original graphene band structures are affected by interlayer coupling. 
	Explicitly, the spectral function is given by   $ A^{l}_{\alpha}(\vec k,\omega)=\sum_{\lambda,\alpha} \left|\braket{\phi_{\vec k,l,\alpha}}{\psi_{\vec k,\lambda}}\right|^2 \delta(\epsilon_{\vec k, \lambda}-\omega)$ where $\ket{\phi_{\vec k,l,\alpha}}$ indicate the original graphene wavefunctions with the layer index  $l=\pm1$ and the sublattice index $\alpha=A, B$. 
	In Fig.~\ref{fig:fig1}(c), we plot the unfolded band structure for the TBGQC approximant,  which around zero energy closely resembles the Dirac cone of the original graphene while significant modification appears in QER.
	
	%%%%%%%%%%%%%%%%%%%%%%%%%%%%%%%%%%%%%%%%%%%%%%%%%%%%%%5
	\begin{figure}[t!]
		\includegraphics[width=\linewidth]{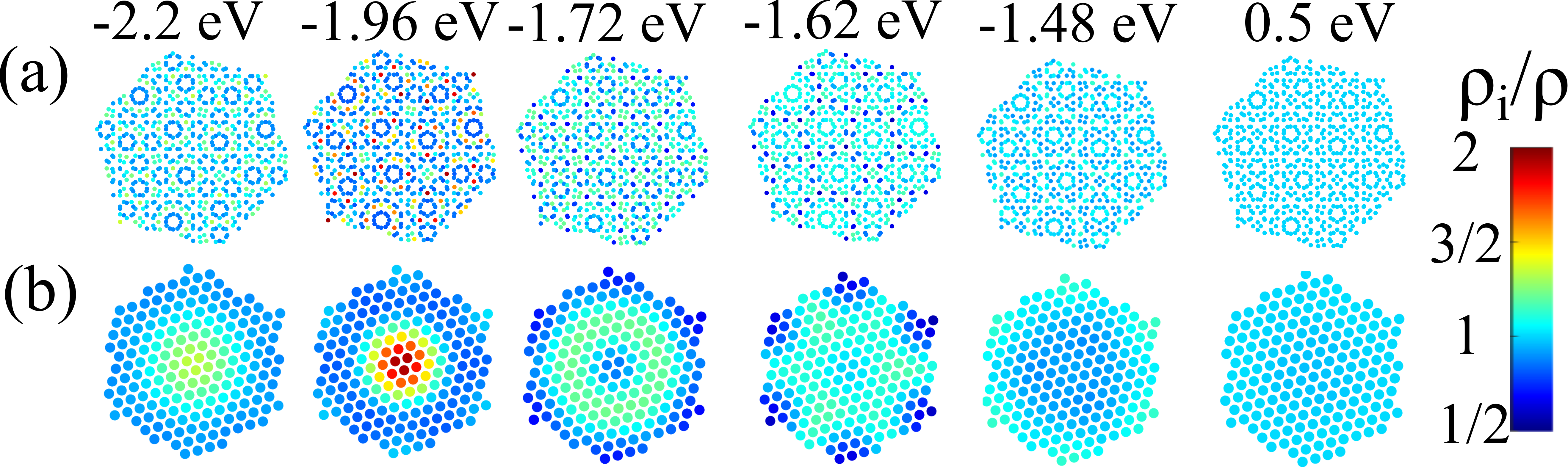}
		\caption{The real space distribution of (a) the LDOS $\rho_i(\omega)$ at different energies $\omega$ (written at the top row).
			(b) LDOS distribution within EBCS for the A sublattices on the top layer projected to the bottom layer (similar plots are obtained for other combinations of the sublattices or layers).  In the plots $\rho_i(\omega)$ is divided by its averages.  The uniform LDOS in a PER ($0.5$ eV) is shown for comparison with other panels in QER.
   }
		\label{fig:fig2}
	\end{figure}
	%%%%%%%%%%%%%%%%%%%%%%%%%%%%%%%%%%%%%%%%%%%%%%%%%%%%%%%%%%%%
	
	In Fig.~\ref{fig:fig2}(a), we plot the LDOS $\rho_i(\omega)$ at different energies $\omega$, 
	which exhibits quasiperiodicity within QER, reflecting the inhomogeneous local environment distribution of lattice sites.
	Contrary to each graphene layer where the lattice sites belonging to a given sublattice (A or B) share the same local environment, 
	the local environments exhibit large fluctuations in TBGQC.  
	To clarify the relation between the local environment and the LDOS inhomogeneity, we introduce an \textit{``environment-based classification space"} (EBCS) as depicted in Fig.~\ref{fig:fig1}(d). 
	To obtain an EBCS, we first project all lattice sites of the top layer belonging to a specific sublattice onto the bottom layer. 
	Then, we record the position of each projected lattice site relative to the hexagonal unit cell of the bottom layer (constructed by connecting nearest neighbor links) enclosing the projected site.  By marking the relative positions of all the projected sites within a single hexagon, we obtain an EBCS (\blue{see the Supporting Information [S3.F]}).
 Because EBCS in TBGQC  contains information about local environments, it resembles the perpendicular space~\cite{DEBRUIJN198139,DEBRUIJN198153} in intrinsic QCs which can classify vertices based on their local environments \cite{PhysRevB.102.115125,PhysRevB.102.224201}. 	
	In Fig.~\ref{fig:fig2}(b), we plot the LDOS distribution in EBCS, which shows smoother variation compared to the original real-space depiction in Fig.~\ref{fig:fig2}(a).
This implies that in QER, vertices with similar local environments yield similar LDOS, thereby reflecting underlying quasiperiodicity, which also alters emergent superconducting instability in TBGQC, as illustrated below.

	To study the superconducting instability of TBGQC, we assume the following phenomenological effective attraction
	\begin{equation}\label{equ:interactiongeneral}
		H_{int}^{\eta}=\sum_{ij}U^{\eta}_{ij}\big(c^\dagger_{i\uparrow}c^\dagger_{j\downarrow}+\eta c^\dagger_{j\uparrow}c^\dagger_{i\downarrow}\big)\big(c_{j\downarrow}c_{i\uparrow}+\eta c_{i\downarrow}c_{j\uparrow}\big),
	\end{equation}
	where $U_{ij}^{\eta}=U_{ji}^{\eta}<0$ is the attraction between electrons at the $i$-th and $j$-th vertices.
 $\eta=+1, -1$ indicate the spin-singlet and spin-triplet pairing channels, respectively.
 % , valid in the absence of spin-orbit coupling   \cite{PhysRevB.99.184514}.
	We investigate the pairing instability using the linearized gap equation approach, which is a powerful tool for studying competing superconducting orders near the critical temperature, and has been extensively applied to various superconductors   \cite{scalapino1986d,RevModPhys.78.17,RevModPhys.83.1589,RevModPhys.84.1383,PhysRevB.99.115122} including the quasicrystalline superconductivity~\cite{PhysRevLett.125.017002}. 
	
	\blue{As shown in the Supporting Information [S2]}, we develop a real-space linearized gap equation suitable for TBGQC approximant given by	$M_{L,L'}\Delta_{L'}^{\eta}= \zeta \Delta_{L}^{\eta}$, where  $M$ is the pairing matrix.
	Here, $L\equiv(i,j)$ represents a pair of vertices that denote the electron positions forming a Cooper pair. 
	To construct the pairing matrix, we consider the pairs of vertices with nonzero attraction $U_{ij}^{\eta}$.
   The order parameter is given by a vector  $\Delta^{\eta}_{L=(i,j)}=U_{ij}^{\eta} \big<c_{j\downarrow}c_{i\uparrow}+\eta c_{i\downarrow}c_{j\uparrow}\big>$  that  contain local pairing information.
	The pairing eigenvalues ($\zeta$) and pairing eigenvectors ($\Delta$) of $M$ give the possible superconducting pairing instabilities of TBGQC.
	As the $\zeta$ increases, one can expect a larger critical temperature for the corresponding pairing channel. 
 Thus,	the dominant pairing instability appears in the channel with the largest $\zeta_{\text{max}}$.
	In deriving the linearized gap equation, we do not assume any particular symmetry or distribution of $\Delta$. This allows us to capture all possible superconducting states including the inhomogeneous pairings such as FFLO~\cite{PhysRev.135.A550} or fractal superconducting states~\cite{Kamiya2018}. 
        Note that FFLO states exhibit spatial modulation on the length scale of the lattice. While these states are generally unfavorable in the absence of a magnetic field ($B$), they can become the dominant pairing in some systems even without $B$, leading to pair-density wave pairings \cite{PhysRevB.94.104508} such as intravalley pairing induced Kekulé ordering \cite{PhysRevResearch.2.043155}.
	We find that the FFLO pairing instabilities (except Kekulé ordering near Dirac point) always exhibit small pairing eigenvalues in our calculations, thus they are neglected in the following discussion (\blue{see the Supporting Information [S2.D]}).
	
	%%%%%%%%%%%%%%%%%%%%%%%%%%%%%%%%%%%%
	\begin{figure}[t!]
		\includegraphics[width=0.95\linewidth]{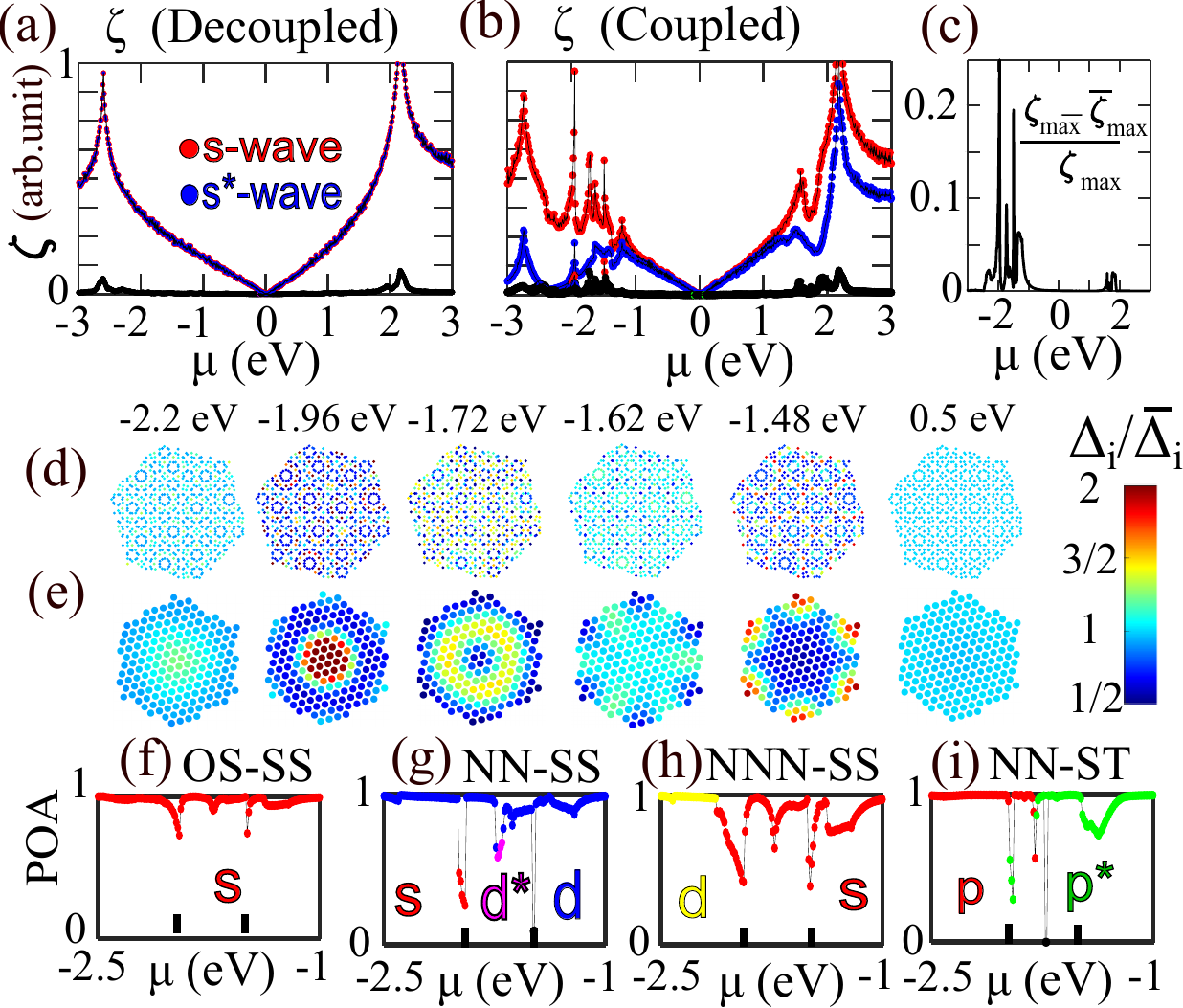}
		\caption{ Superconductivity in TBGQC. 
			(a, b) The dominant pairing channels with the largest eigenvalues of $M$ for onsite spin singlet attraction as a function of the chemical potential $\mu$ for (a) layer decoupled and (b) layer coupled cases.
			(a) The red and blue lines indicate the leading s-wave and s$^*$-wave pairings defined in the layer decoupled limit, respectively, while the black indicates the subleading pairing eigenvalue with lower symmetry (such as FFLO instabilities).
			In (b), the red and blue colors indicate the largest and second-largest pairing channels which are nearly identical to the s-wave and s$^*$-wave pairings, respectively.
			(c) The comparison between the largest eigenvalue of $M$ and $\overline{M}$ (denoted by $\zeta_{\text{max}}$, $\overline\zeta_{\text{max}}$, respectively) for onsite spin singlet attraction, which shows relative enhancement in QER.
			(d, e) Distribution of dominant pairing eigenvector $\Delta_i=\Delta^{+1}_{(i,i)}$  for the onsite attraction at a different chemical potential (written on the top of panels) in (d) real space and (e) EBCS. 
			In (d,e), $\Delta_i$ is divided by its averages $\bar{\Delta_i}$. 
			(f-i) Pairing overlap amplitude (POA) of the dominant pairing channels and it's characteristic for: (f) onsite [OS] spin-singlet [SS], (g) nearest-neighbor [NN] SS,  (h) next-nearest-neighbor [NNN] SS, and (i) NN spin-triplet [ST] attraction. Smaller POA indicates a non-uniform distribution of pairing amplitudes.
   } 
		\label{fig:fig3}
	\end{figure}
	%%%%%%%%%%%%%%%%%%%%%%%%%%%%%%%%%%%%%%%

	Let's first consider the onsite attraction represented by $U_{i,j}=-\delta_{i,j}$, which exclusively results in spin-singlet pairings~\cite{PhysRevLett.98.146801,PhysRevB.99.184514}. 
	As higher DOS leads to higher superconducting transition temperatures in BCS theory~\cite{PhysRev.106.162},
	the LDOS enhancement due to quasiperiodicity may augment the quasiperiodic superconducting instabilities in TBGQC~\cite{PhysRevB.98.184510,PhysRevB.107.174508,PhysRevResearch.3.023195}. 
	To confirm it, we consider the largest eigenvalues of the pairing matrix for decoupled and coupled graphene layers in Fig.~\ref{fig:fig3}(a) and (b), respectively.
	One can observe that superconductivity is indeed enhanced in QER when two layers are coupled.  
	To demonstrate the relation between the LDOS enhancement and enhanced superconductivity, we compare the largest eigenvalues of $M$ and $\overline{M}$ (denoted by $\zeta_{\text{max}}$, $\overline{\zeta}_{\text{max}}$, respectively).
	Here  $\overline{M}_{l,l'}=\tfrac{1}{N_{\text{cell}}}\sum_{L\in l; L'\in l'} M_{L,L'}$  is a $2\times2$ pairing matrix obtained by averaging the matrix elements of $M$ with identical layer indices ($l,l'=\pm$),  \blue{see the Supporting Information [S1]}.
	Therefore, $\overline{M}$ describes uniform superconductivity within each layer, thus benefiting only from DOS but not LDOS. 
	In Fig.~\ref{fig:fig3}(c), we plot $(\zeta_\text{max}-\overline\zeta_\text{max})/\zeta_\text{max}$ which shows relative enhancement of the dominant pairing eigenvalue.
	Notably, the pairing eigenvalue exhibits a significant enhancement within QER. In contrast, no such enhancement is observed in other energy ranges.

	To systematically examine the dominant pairing profile, we compare $\Delta$ with trial pairing eigenvectors ($\Delta^t_{\alpha}$, where $\alpha$ labels different trial states such as s-wave, etc.)  with specific pairing symmetry defined in the layer decoupled limit.
	We compute the pairing overlap amplitude (POA) of $\Delta$ with the trial pairing eigenvectors. The POA is calculated as $|\Delta \cdot \Delta^t_{\alpha}|^2$, and we sum the POAs of trial eigenvectors with the same symmetry (\blue{see the Supporting Information [S3]}). If POA approaches 1 (0), the trial pairing eigenvector can (cannot) fully describe the given pairing eigenvector.
   Generally, POA smaller than 1 indicates the presence of non-uniform and aperiodic pairing.
	We denote the characteristic of the given pairing eigenvector using the symmetry of the trial pairing eigenvector with the largest POA.
	In Fig.~\ref{fig:fig3}(a) and (b), we characterize the pairing instabilities with red and blue colors indicating the s-wave and s$^*$-wave symmetries [obtained in the layer decoupled limit, respectively].
	Here and in the following, the pairing symmetry with (without) $^*$ symbol indicates that the pairing eigenvector has the same (opposite) signs between layers.
	As shown in Fig.~\ref{fig:fig3}(a), both s-wave and s$^*$-wave symmetries are degenerate in the layer decoupled limit. 
	On the other hand, as demonstrated in Fig.~\ref{fig:fig3}(b), when two graphene layers are coupled, $\Delta$ with s-wave characteristic dominates over s$^*$-wave pairing.
	As shown in Fig.~\ref{fig:fig3}(f),  the POA of the dominant $\Delta$ deviates significantly away from the s-wave pairing in QER
 while in other energy ranges, POA is one, indicating uniform s-wave pairing in both layers. 
	We note that increasing approximant generation toward quasiperiodic structure further enhances the quasiperiodic pairing instabilities (\blue{see the Supporting Information [Fig. S10]}).

  In Fig.~\ref{fig:fig3}(d) and (e), we depict the distribution of the dominant pairing eigenvector within a supercell and an EBCS, respectively.
	The dominant $\Delta$ does not change its sign across the entire system, which, in line with Fig.~\ref{fig:fig3}(b), can be considered as an inhomogeneous s-wave pairing.
	Comparing Fig.~\ref{fig:fig2}(a,b) with Fig.~\ref{fig:fig3}(d,e), it becomes apparent that  $\Delta$ mirrors the underlying LDOS distribution. 
	Similar to LDOS distribution, $\Delta$ also exhibits quasiperiodicity within QER, while it is uniform at other energy ranges (see e.g. Fig.~\ref{fig:fig3} (d,e) for $\omega=0.5 \text{eV}$ ).
	The order parameter and LDOS are distributed around their average values, remaining nonzero at all vertices, which resembles the characteristics observed in weakly disordered superconductors~\cite{PhysRevLett.81.3940} but is distinct from highly disordered superconductors with superconducting islands~\cite{Dubi2007}.
	The absence of superconducting islands despite the underlying quasiperiodicity indicates that superconductivity remains coherent and is not suppressed by phase fluctuations~\cite{PhysRevResearch.3.023195}.

	Furthermore, we consider the effect of longer-range [nearest-neighbor (NN) and next-nearest-neighbor (NNN)] attractive interactions in spin-singlet (SS) and spin-triplet (ST) channels \cite{PhysRevB.108.134514} (see further detail in \blue{Supporting Information [S3]}).  
 For monolayer graphene, due to the underlying $D_{6h}$ symmetry, p-wave ($p_x$, and $p_y$) and d-wave ($d_{xy}$, and $d_{x^2-y^2}$) pairings are doubly degenerate, while f-wave and s-wave pairings are not \cite{Black-Schaffer_2014}. In TBGQC, due to the $D_{6d}$ symmetry \cite{PhysRevB.99.165430}, f-wave pairing  also becomes doubly degenerate ($f_{y^3-3yx^2}$, and $f_{x^3-3xy^2}$), similar to p-wave ($p_x$, and $p_y$),  p*-wave ($p^*_x$, and $p^*_y$), d-wave ($d_{xy}$, and $d_{x^2-y^2}$) and d*-wave ($d^*_{xy}$, and $d^*_{x^2-y^2}$) pairings.

	 In the NN-SS attraction, mostly d-wave and d\(^*\)-wave pairing instabilities are competing. However, within QER d-wave pairing becomes dominant pairing instability, but its POA becomes smaller than 1 [Fig.~\ref{fig:fig3}(g)].
		Especially, highly quasiperiodic pairing with small POA and mostly s-wave characteristic emerges at around $\mu=-1.96$ eV [see red dots in Fig.~\ref{fig:fig3}(g)]. 
		In the case of the  NNN-SS attraction, mostly s-wave pairing is dominant, although close to the main vHS $\mu\lessapprox -2.2$ eV, d-wave becomes dominant.  
		As expected, within QER [see Fig.~\ref{fig:fig3}(h)],  POA is less than one, indicating underlying quasiperiodic pairing.  

	For the NN-ST attraction,  the dominant instability is p-wave or p$^*$-wave pairing. However, near zero energy, four-fold degenerate Kekule ordering~\cite{PhysRevB.82.035429,Zhou_2013} dominates \blue{(see Supporting Information [Fig. S4])}. Like other channels, POA of dominant pairings (p-wave or p*-wave) drops away from 1 in QER [Fig.~\ref{fig:fig3} (i)]. Finally, in the NNN-ST attraction, f-wave pairings dominate and show no significant aperiodicity even within QER~\cite{Zhang2015,PhysRevB.107.224511}, which may arise from maximizing f-wave pairing eigenvalue in QER as shown in \blue{the Supporting Information [S3.D]} using perturbation theory.

To conclude, we have demonstrated the appearance of quasiperiodic pairing instabilities with s-wave characteristics in QER for different spin-singlet attraction channels, especially near QCvHS. 
This suggests that quasiperiodic s-wave pairing instability is generally expected to be dominant in TBGQC when the Fermi level is tuned to reach QER.
We note that electrostatic gating may not be sufficient to dope enough carriers to reach the QER. However, external chemical doping might be used to largely tune the Fermi level, as demonstrated in various two-dimensional materials \cite{PhysRevLett.104.136803,PhysRevLett.125.176403,PhysRevMaterials.3.110301,doi:10.1126/science.aaa6486,Kanahashi2019,9173729}. 

Moreover, our theory can also be applied to the superconductivity in the recently discovered moir\'e graphene quasicrystal constructed by stacking three layers with small rotation.
Interestingly, moir\'e graphene quasicrystal shows both features of moir\'e energy range and QER in which superconductivity emerges~\cite{Uri2023}.
In the spectral function of the moir\'e quasicrystal, akin to TBGQC \cite{PhysRevB.99.165430}, many weakly dispersing states (partial flat bands) are observed within QER\cite{Uri2023}.
We speculate that the large LDOS fluctuations induced by quasiperiodicity, and the existence of partial flat bands, may be a crucial driving force behind the emergent superconductivity in moiré quasicrystals.

 \section*{Acknowledgements}
	\begin{acknowledgements}
     R.G. thanks Mikito Koshino for his insightful discussion. 
    R.G. thanks SungBin Lee for hosting R.G. in KAIST and insightful discussion.  
    R.G. and B.J.Y. were supported by 
Samsung Science and Technology Foundation under Project No.
 SSTF-BA2002-06, National Research Foundation of Korea
 (NRF) grants funded by the government of Korea (MSIT)
 (Grants No. NRF-2021R1A5A1032996), and GRDC(Global
 Research Development Center) Cooperative Hub Program
 through the National Research Foundation of Korea(NRF)
 funded by the Ministry of Science and ICT(MSIT) (RS-2023
00258359).
	\end{acknowledgements}

	\section*{Supporting Information}
	See \blue{Supporting Information} for \blue{(S1)} the tight-binding model, \blue{(S2)} derivation of the linearized gap equation, \blue{(S3-S4)} discussion on FFLO states, phase diagrams for different attraction channels, trial pairing eigenvectors and pairing overlap amplitude, aperiodicity degree for f-wave pairing, environment based classification space, and further discussions or figures. 
	\section*{References} 
	\bibliography{refs}
	
	\clearpage
        \newpage
        \begin{figure}
            \centering
	\includegraphics[page=1, width=\textwidth]{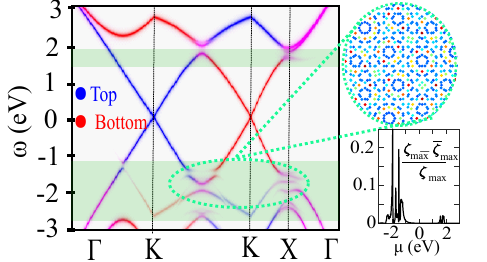}
	\caption{This is the TOC graphic}
            \label{fig:TOCgraphic.}
        \end{figure}

\end{document}

%	%%%%%%%%%%%%%TITLE%%%%%%%%%%%%%
	\title{Supplemental Material for "Quasiperiodic pairing in graphene quasicrystals"}
	%%%%%%%%%%%%%%%%%%%%%%%%%%%%%%%
	
	%%%%%%%%%%%%AUTHORS%%%%%%%%%%%%
	\author{Rasoul Ghadimi}
	% \orcidlink{0000-0001-6927-4008}
	\affiliation{Department of Physics and Astronomy, Seoul National University, Seoul 08826, Korea}
	\affiliation{Center for Theoretical Physics (CTP), Seoul National University, Seoul 08826, Korea}
	\affiliation{ Institute of Applied Physics, Seoul National University, Seoul 08826, Korea}

	\author{Bohm-Jung Yang}
	\email{bjyang@snu.ac.kr}
	\affiliation{Department of Physics and Astronomy, Seoul National University, Seoul 08826, Korea}
	\affiliation{Center for Theoretical Physics (CTP), Seoul National University, Seoul 08826, Korea}
	\affiliation{ Institute of Applied Physics, Seoul National University, Seoul 08826, Korea}

	\maketitle
	\tableofcontents
	%%%%%%%%%%%%%BACKGROUND%%%%%%%%%%%%%%%%%%

\setcounter{figure}{0}
\renewcommand{\thefigure}{S\arabic{figure}}
\renewcommand{\thesection}{S\arabic{section}}
\renewcommand{\theequation}{S\arabic{equation}}

\section{Tight-binding calculation}
In this work, we use the conventional \( p_z \) orbital-based tight-binding Hamiltonian to describe the carbon atoms in a graphene quasicrystal \cite{PhysRevB.102.115123,Yu2019}. The Hamiltonian is given by:

\begin{equation}
    H_0=\sum_{i,j,\sigma} t(\vec{r_i}-\vec{r_j})c^\dagger_{i\sigma}c_{j\sigma}-\mu \sum_{i,\sigma} c^\dagger_{i\sigma}c_{i\sigma},
\end{equation}

where \( c^\dagger_{i\sigma} \) and \( c_{i\sigma} \) are the creation and annihilation operators of an electron with spin \(\sigma = \uparrow, \downarrow \) at the \( i \)-th site located at position \( \vec{r_i} \). The parameter \(\mu\) represents the chemical potential. In this Hamiltonian, the hopping energy between vertices is given by:

\begin{equation}
    t(\vec{r}) = \gamma_1 \frac{z^2}{r^2}e^{2.218(h_z-|r|)} - \gamma_0\left(1 - \frac{z^2}{r^2}\right) e^{2.218(b-|r|)},
\end{equation}

where \( h_z = 3.349 \, \text{\AA} \) and \( b = 1.418 \, \text{\AA} \) are the interlayer distance and the shortest carbon-carbon distance, respectively. The Slater-Koster parameters, obtained from \cite{doi:10.1126/science.aar8412}, are \(\gamma_0 = 2.7 \, \text{eV} \) and \(\gamma_1 = 0.48 \, \text{eV} \). 
In our calculations, we consider hopping interactions within a distance less than \( r_c \equiv 3.1b \approx 4.4 \, \text{\AA} \), which is sufficient to capture the essential physics of the graphene quasicrystal.

\section{ Linearized gap equation}
\label{Supp:LGE}
\subsection{Attraction, mean field and  Linearized gap equation}
This section explain our mean-field method for investigating potential superconducting pairing instabilities. Here, we focus on the phenomenological attraction between electrons, either in singlet or triplet form. The triplet \([ \eta=+1 ]\) and singlet \([ \eta=-1 ]\) interactions are described by:

\begin{equation}\label{interactiongeneral}
	H_{int}^{\eta}=\sum_{ij}U^{\eta}_{ij}\big(c^\dagger_{i\uparrow}c^\dagger_{j\downarrow}+\eta c^\dagger_{j\uparrow}c^\dagger_{i\downarrow}\big)\big(c_{j\downarrow}c_{i\uparrow}+\eta c_{i\downarrow}c_{j\uparrow}\big),
\end{equation}

where \( U_{ij}^{\eta}=U_{ji}^{\eta} \). In Eq.~(\ref{interactiongeneral}), \( U_{ij}^{\eta} \) represents the singlet or triplet attraction. By applying the real-space mean-field (MF) approximation \([ AB \approx \langle A \rangle B + A \langle B \rangle - \langle A \rangle \langle B \rangle ]\) to Eq.~(\ref{interactiongeneral}), we obtain:

\begin{equation}\label{MFHAMILTONIAN}
	H_{MF}=\sum_{i,j}\sum_{\eta=\pm1} \Delta^{\eta}_{ij}\big(c^\dagger_{i\uparrow}c^\dagger_{j\downarrow}+\eta c^\dagger_{j\uparrow}c^\dagger_{i\downarrow}\big)+H.c.-\sum_{ij,\eta=\pm1}\frac{|\Delta^{\eta}_{ij}|^2}{U_{ij}},
\end{equation}

where \( \Delta^{\eta}_{ij}=\pm\Delta^{\eta}_{ji} \) denotes the singlet \([ \eta=1 ]\) and triplet \([ \eta=-1 ]\) pairing potentials or order parameters between electrons located at sites \( i \) and \( j \). These pairing potentials are determined by the self-consistency equation:

\begin{equation}\label{pairingpotential}
	\Delta^{\eta}_{ij}=U_{ij}^{\eta} \big\langle c_{j\downarrow}c_{i\uparrow}+\eta c_{i\downarrow}c_{j\uparrow} \big\rangle.
\end{equation}

Note that we consider only spin-triplets along the \( \hat{z} \)-direction; other triplet instabilities can be obtained via spin rotation.

Order parameters within our MF approximation are derived using the self-consistency equation given in Eq.~(\ref{pairingpotential}). This approach allows us to determine the critical temperature \( T_c \) and pairing amplitude for various parameters such as filling temperature and interaction profile. However, in self-consistency equations, the solution typically converges to the most probable pairing based on an initial configuration for \( \Delta^{\eta}_{ij} \), making it challenging to identify the competition between different pairing instabilities.

To explore competing channels near the critical temperature, we can reformulate Eq.~(\ref{pairingpotential}) as an eigenvalue problem, known as the linearized gap equation. Assuming \( T \approx T_c \), we expect small pairing potentials, allowing us to find the expectation value in Eq.~(\ref{pairingpotential}) perturbatively.

To obtain the real space linearized gap equation, we start with diagonalizing tight-binding Hamiltonian  using 
\begin{equation}\label{transformation}
	c_{i\sigma}=\sum_{\alpha}\overline{\psi^i_{\alpha}}c_{\alpha\sigma},\qquad \text{and} \qquad
	c^\dagger_{i\sigma}=\sum_{\alpha}\psi^i_{\alpha}c^\dagger_{\alpha\sigma},
\end{equation}
where bars mean complex conjugate and $\alpha$ is the energy index. In Eq.~(\ref{transformation}) and the following, we use Greek and Latin letters for energy and site index, receptively.  
The total MF Hamiltonian, including chemical potential (which controls filling), reads,
\begin{equation}\label{MFinnormalbasis}
 \begin{split}
	&H_{Sc}=H_{tb}+\lambda H_{MF}+E_0,\\
	&H_{0}=\sum_{\alpha}\tilde\epsilon_{\alpha} \big|e,\alpha,\uparrow\big>\big<e,\alpha,\uparrow\big|-\tilde\epsilon_{\alpha} \big|h,\alpha,\downarrow\big>\big<h,\alpha,\downarrow\big|,\\
	&H_{MF}=\sum_{\alpha,\beta}\Delta_{\alpha\beta}
	\big|e,\alpha,\uparrow\big>\big<h,\beta,\downarrow\big|+H.c,
 \end{split}
\end{equation}
where e and h refer to electron and hole sectors of Bogoliubov-de-Gennes (BdG) Hamiltonian. In Eq.~(\ref{MFinnormalbasis}), $\tilde\epsilon_{\alpha}=\epsilon_{\alpha}-\mu$, $E_0=-\sum_{ij,\eta=\pm1}{|\Delta^{\eta}_{ij}|^2}\big/{U_{ij}}+\sum_{\alpha}\epsilon_{\alpha}$ and $\Delta_{\alpha\beta}$ is given by
\begin{equation}\label{orderparameterbasismain}
	\Delta_{\alpha\beta}=\sum_{\eta=\pm}\sum_{i,j} \Delta^{\eta}_{ij}
	\big(	\psi^i_{\alpha}\psi^j_{\beta}+\eta \psi^j_{\alpha}\psi^i_{\beta}		\big).
\end{equation}
In the following, we take $\lambda$ for perturbation theory and finally, we put it $\lambda=1$.
In Eq.~(\ref{MFinnormalbasis}), we use the following notation interchangeably
\begin{equation}\label{braketnotation}
\begin{split}
  &  	\big|e,\alpha,\uparrow\big>=c^\dagger_{\alpha\uparrow},\quad 	\big<e,\alpha,\uparrow\big|=c_{\alpha\uparrow},\\
&	\big|h,\alpha,\downarrow\big>=c_{\alpha\downarrow},\quad 	\big<h,\alpha,\downarrow\big|=c^\dagger_{\alpha\downarrow},
\end{split}
\end{equation}
and in terms of these basis, pairing potential, Eq.~(\ref{pairingpotential}) is given by
\begin{equation}\label{pairingpotentialnewbasis}
	\Delta^{\eta}_{ij}=\sum_{\alpha,\beta}U_{ij}^{\eta} 
	\overline{\big(
		\psi^j_{\alpha}\psi^i_{\beta}+\eta 	\psi^i_{\alpha}\psi^j_{\beta} 
		\big)}
	\Big<
	\big|h,\alpha,\downarrow\big>\big<e,\beta,\uparrow\big|
	\Big>.
\end{equation}
As we mentioned before, close to $T\approx T_c$ order parameters are infinitesimal, and we can find the expectation value in Eq.~(\ref{pairingpotentialnewbasis}) perturbatively. In perturbation theory, energy and eigenvector of perturbed Hamiltonian is given by
\begin{equation}\label{perturbationtheory}
\begin{split}
&	\tilde\epsilon_{\alpha,e}=\sum_{n=0}\lambda^nE^n_{e,\alpha},\\
&\ket{e,\alpha,\uparrow}=\sum_{n=0} \lambda^nC^n_{ee,\alpha,\beta} \ket{e,\beta,\uparrow}+\sum_{n=0} \lambda^nC^n_{eh,\alpha,\beta} \ket{h,\beta,\downarrow},\\
&	\tilde\epsilon_{\alpha,h}=\sum_{n=0}\lambda^nE^n_{h,\alpha},\\ 
&\ket{h,\alpha,\downarrow}=\sum_{n=0} \lambda^nC^n_{he,\alpha,\beta} \ket{e,\beta,\uparrow}+\sum_{n=0} \lambda^nC^n_{hh,\alpha,\beta} \ket{h,\beta,\downarrow},    
\end{split}
\end{equation}
where,
\begin{equation}\label{perturbationtheory2}
\begin{split}
	E^0_{e,\alpha}=\tilde\epsilon_{\alpha},\qquad  C^0_{ee,\alpha,\beta}=\delta_{\alpha,\beta},\qquad C^0_{eh,\alpha,\beta}=0,\\
	E^0_{h,\alpha}=-\tilde\epsilon_{\alpha},\qquad  C^0_{he,\alpha,\beta}=0,\qquad C^0_{hh,\alpha,\beta}=\delta_{\alpha,\beta},    
\end{split}
\end{equation}
and we obtain the remaining coefficient, perturbatively using
\begin{eqnarray}\label{perturbationtheory3}
\begin{split}
    &	E^n_{e,\alpha}=\sum_{\beta} C^{n-1}_{eh,\alpha,\beta} \Delta_{\alpha\beta},\\
&	E^n_{h,\alpha}=\sum_{\beta} C^{n-1}_{he,\alpha,\beta} \Delta^*_{\beta\alpha},\\
&	C^{n}_{ee,\alpha\beta}=\frac{1}{\tilde\epsilon_{\alpha}-\tilde\epsilon_{\beta}}\left[\sum_{\gamma}C^{n-1}_{eh,\alpha,\gamma}\Delta_{\beta\gamma}-\sum_{n_1=1}^{n-1} E^{n_1}_{e,\alpha}C^{n-n_1}_{ee,\alpha\beta}\right]\\
&	C^{n}_{eh,\alpha\beta}=\frac{1}{\tilde\epsilon_{\alpha}+\tilde\epsilon_{\beta}}\left[\sum_{\gamma}C^{n-1}_{ee,\alpha,\gamma}\Delta^*_{\gamma\beta}-\sum_{n_1=1}^{n-1} E^{n_1}_{e,\alpha}C^{n-n_1}_{eh,\alpha\beta}\right]\\	
&	C^{n}_{he,\alpha\beta}=\frac{1}{-\tilde\epsilon_{\alpha}-\tilde\epsilon_{\beta}}\left[\sum_{\gamma}C^{n-1}_{hh,\alpha,\gamma}\Delta_{\beta\gamma}-\sum_{n_1=1}^{n-1} E^{n_1}_{h,\alpha}C^{n-n_1}_{he,\alpha\beta}\right]\\
&	C^{n}_{hh,\alpha\beta}=\frac{1}{-\tilde\epsilon_{\alpha}+\tilde\epsilon_{\beta}}\left[\sum_{\gamma}C^{n-1}_{he,\alpha,\gamma}\Delta^*_{\gamma\beta}-\sum_{n_1=1}^{n-1} E^{n_1}_{h,\alpha}C^{n-n_1}_{hh,\alpha\beta}\right]
\end{split}		
\end{eqnarray}
It is easy to show that only odd [even] integer for $C^{n}_{eh,\alpha\beta}$, and $C^{n}_{he,\alpha\beta}$ [$C^{n}_{ee,\alpha\beta}$, $C^{n}_{hh,\alpha\beta}$, $E^n_{e,\alpha}$, and $E^n_{h,\alpha}$] will survive in Eq.~(\ref{perturbationtheory}). 
In following, we need only find eigenvectors of Eq.(\ref{MFinnormalbasis}) using the first-order perturbation theory and we get
\begin{equation}\label{eigenstateperturbation}
	\big|e,\beta,\uparrow\big>_{sc}\approx\big|e,\beta,\uparrow\big>+
	\sum_{\alpha}\frac{\Delta^*_{\beta\alpha}}{\tilde\epsilon_{\beta}-(-\tilde\epsilon_{\alpha})}\big|h,\alpha,\downarrow\big>,
 \end{equation}
 and
 \begin{equation}
	\\\label{eigenstateperturbation1}
	\big|h,\alpha,\downarrow\big>_{sc}\approx\big|h,\alpha,\downarrow\big>+
	\sum_{\beta}\frac{\Delta_{\beta\alpha}}{(-\tilde\epsilon_{\alpha})-\tilde\epsilon_{\beta}}\big|e,\beta,\uparrow\big>.
\end{equation}
By putting these eigenvectors in Eq.~(\ref{pairingpotentialnewbasis}),  and defining $F[\epsilon]=1/(1+\exp\beta\epsilon)$,  $\beta=1/T$ we get
\begin{equation}\label{pairingpotentialsubsitut1}
	\Delta^{\eta}_{ij}=\sum_{\alpha,\beta}U_{ij}
	\overline{ \big(	\psi^j_{\alpha}\psi^i_{\beta}+\eta \psi^i_{\alpha}\psi^j_{\beta}  \big)}
	\Big(
	\frac{	F[\tilde\epsilon_{\beta}]-F[-\tilde\epsilon_{\alpha}]}{\tilde\epsilon_{\beta}+\tilde\epsilon_{\alpha}}
	\Big)\Delta_{\beta\alpha}.
\end{equation}
By substituting $\Delta_{\alpha\beta}$ from Eq.~(\ref{orderparameterbasismain}) and considering $F[-\epsilon]=1-F[\epsilon]$ we find
\begin{equation}\label{linearized1}
\begin{split}
   & 	\Delta^{\eta}_{ij}=\\
  & \sum_{\alpha,\beta}U_{ij} 
	\overline{\big(\psi^j_{\alpha}\psi^i_{\beta} +\eta \psi^i_{\alpha}\psi^j_{\beta}	\big)}
	\Big(
	\frac{	F[\tilde\epsilon_{\beta}]+F[\tilde\epsilon_{\alpha}]-1}{\tilde\epsilon_{\beta}+\tilde\epsilon_{\alpha}}
	\Big)\times\\
 &
	\sum_{\eta'=\pm}\sum_{k,l} \Delta^{\eta'}_{kl}
	\big(	\psi^k_{\beta}\psi^l_{\alpha}+\eta \psi^l_{\beta}\psi^k_{\alpha}		\big).
\end{split}
\end{equation}
Finally, we can show  Eq.~(\ref{linearized1}) by matrix multiplication,
\begin{equation}\label{matrixlienarized}
	\Delta^{\eta}_{ij}=\sum_{k,l}M_{ij,kl}\Delta^{\eta}_{kl},
\end{equation}
where
\begin{equation}\label{Mmatrix}
\begin{split}
&	M_{ij,kl}=\\
&U_{ij}^{\eta}   \sum_{\alpha,\beta}	\frac{	F[\tilde\epsilon_{\beta}]+F[\tilde\epsilon_{\alpha}]-1}{\tilde\epsilon_{\beta}+\tilde\epsilon_{\alpha}}	\overline{\big(\psi^j_{\alpha}\psi^i_{\beta} +\eta \psi^i_{\alpha}\psi^j_{\beta}	\big)}
	\big(	\psi^k_{\beta}\psi^l_{\alpha}+	\eta\psi^k_{\alpha}\psi^l_{\beta}		\big).    
\end{split}
\end{equation}
Therefore, the original self-consistency equation is mapped to find the critical temperature, where the largest pairing eigenvalue of $M$ becomes precisely one. The corresponding eigenvector gives the profile of pairing potential near the phase transition. We can simplify calculations by choosing a temperature in Eq.~(\ref{Mmatrix}) and anticipating its eigenvectors similar to those near the critical temperature. By this approximation, we can expect that other eigenvectors with smaller pairing eigenvalues are subdominant pairing channels. Therefore, we can study competition between different pairing channels by comparing their corresponding pairing eigenvalue. 
We found that changing temperature in the calculation, do not change results (see Fig.~\ref{fig:fig5}).

\subsection{Uniform pairing}

 In the following, we want to find how to assume uniform pairing through the lattice and then use benefit of super-cell momentum space.
 In this sense, we first classify pairs of indices that we expect to give similar pairing potential and label them with their class, $\mathfrak{l}$, and its address in that class, $\mathfrak{s}$, or $ij\leftrightarrow \mathfrak{l,s}$.
Therefore, in this new notation, we can show Eq.~(\ref{matrixlienarized}) with  
\begin{equation}\label{identifypairingpotentital0}
	\Delta^{\eta}_{ij}\equiv	\Delta^{\eta}_{\mathfrak{ls}}=\sum_{\mathfrak{l's'}}M_{\mathfrak{ls},\mathfrak{l's'}}\Delta^{\eta}_{\mathfrak{l's'}}.
\end{equation}
By identifying pairing potentials within each classes, $\Delta^{\eta}_{\mathfrak{ls}}=\Delta^{\eta}_{\mathfrak{l}}$  we get
\begin{equation}\label{identifypairingpotentital1}
	\Delta^{\eta}_{\mathfrak{l}}=\sum_{\mathfrak{l'}} M_{\mathfrak{ll'}}	\Delta^{\eta}_{\mathfrak{l'}}.
\end{equation}
In Eq.~(\ref{identifypairingpotentital1}), the $M_{ll'}$ has a smaller dimension and given by
\begin{equation}\label{identifyM}
	\begin{split}
	   & M_{\mathfrak{ll'}}=\\
    &\frac{1}{n_{\mathfrak{l}}}
	\sum_{	ij\in\mathfrak{l},	kl\in\mathfrak{l'}}
	U_{ij}^{\eta}   \sum_{\alpha,\beta}	\frac{	F[\tilde\epsilon_{\beta}]+F[\tilde\epsilon_{\alpha}]-1}{\tilde\epsilon_{\beta}+\tilde\epsilon_{\alpha}} \times\\
 &	\overline{\big(\psi^j_{\alpha}\psi^i_{\beta} +\eta \psi^i_{\alpha}\psi^j_{\beta}	\big)}
	\big(	\psi^k_{\beta}\psi^l_{\alpha}+	\eta\psi^k_{\alpha}\psi^l_{\beta}		\big),
	\end{split}
\end{equation}
where $n_{\mathfrak{l}}$ is the number of element in the given class.

\subsection{Benefiting from approximant translational symmetry}
If the systems respect approximate translational symmetry, we can simplify the calculations even further. This is important because, although a large supercell can exhibit aperiodic pairing, choosing a very large supercell is impractical due to the high computational cost of matrix diagonalization. Therefore, one can utilize a sufficiently large approximant and benefit from its translational symmetry. This approach also enables us to access more states near the Fermi energy, leading to more precise calculations. Using the spectral function, the usual Fermi surface in the decoupled limit can be captured only when the system is large enough. In our calculations, we consider systems that effectively contain a few million carbon atoms (50 by 50 momentum mesh).  In following we explain how to use supercell momentum space in the linearized gap equation. 

To start suppose that we make a system which contains several approximants [or unit-cells]. 
If we assume that the pairing potential respects the approximant's translational symmetry,  we can identify all pairings whose corresponding vertices are repeated in the system. 
Furthermore, due to the translational symmetry, we can express the TBG Hamiltonian in momentum space using the Fourier transformation of fermionic operators, 
\begin{equation}\label{fouriertransformation}
	\begin{split}
	    c^{\dagger}_{i\sigma}=\frac{1}{\sqrt{N}}\sum_{\vec{k}\in S.C.B}e^{i\vec{k}.(\vec{R_i}+\vec{r}_i)} c^{\dagger}_{\mathrm{s}_i,\vec{k},\sigma},\\ 
	c^{\dagger}_{\mathrm{s}_i,\vec{k},\sigma}=\frac{1}{\sqrt{N}}\sum_{ i \in s_i}e^{-i\vec{k}.(\vec{R_i}+\vec{r}_i)} 	c^{\dagger}_{i\sigma},
	\end{split}
\end{equation}
where $i$ is defined by its unit cell index, lattice vector, and sub-super-lattice position, $\mathrm{s}_i$, $\vec{R_i}$, and $\vec{r_i}$, respectively. In Eq.~(\ref{fouriertransformation})
S.C.B is the super-cell Brillouin zone, and $N$ is the number of the approximant.

The TBG Hamiltonian in Momentum space is given by
\begin{equation}\label{hamiltoninamomentumspace}
	H=\sum_{\vec{k}\in B_{S.C}}\sum_{i,j} t(\vec{r_i}-\vec{r_j})e^{i \vec{k}.(R_i+r_i-R_j-r_j)} c^\dagger_{\mathrm{s}_i\vec{k}\sigma}c_{\mathrm{s}_j\vec{k}\sigma}.
\end{equation}
We can diagonalize Eq.~(\ref{hamiltoninamomentumspace}) using
\begin{equation}\label{diagonalizehamilkspace}
	\begin{split}
	    c_{s_i,\vec{k},\sigma} =\sum_{\alpha}\overline{\psi^{\mathrm{s}_i}_{\alpha,\vec{k}}} c_{\alpha,\vec{k},\sigma}
	,\\
	c^{\dagger}_{s_i,\vec{k},\sigma} =\sum_{\alpha}\psi^{\mathrm{s}_i}_{\alpha,\vec{k}} c^{\dagger}_{\alpha,\vec{k},\sigma}.
	\end{split}
\end{equation}
By combine Eq.~(\ref{diagonalizehamilkspace}) with Eq.~(\ref{fouriertransformation}),
\begin{equation}\label{diagonalizehamilkspace2}
\begin{split}
    	c^{\dagger}_{i\sigma}=\frac{1}{\sqrt{N}}\sum_{\alpha,\vec{k}\in S.C.B}e^{i\vec{k}.(\vec{R_i}+\vec{r}_i)} \psi^{\mathrm{s}_i}_{\alpha,\vec{k}} c^{\dagger}_{\alpha,\vec{k},\sigma},
	\\
	c_{i\sigma}=\frac{1}{\sqrt{N}}\sum_{\alpha,\vec{k}\in S.C.B}e^{-i\vec{k}.(\vec{R_i}+\vec{r}_i)} \overline{\psi^{\mathrm{s}_i}_{\alpha,\vec{k}}} c_{\alpha,\vec{k},\sigma},
\end{split}
\end{equation}
where $\alpha$ is the band index. 
By comparing Eq.~(\ref{diagonalizehamilkspace2}) with Eq.~(\ref{transformation}), and identifying pairing between two vertices $i,j$ with the same $s_i,s_j$, we can  read Eq.~(\ref{identifyM}) 
\begin{equation}
\label{Mkspacemodemnt}        
\begin{split}
    	M_{\mathrm{s_is_js_ls_m}}&=\frac{1}{N^3}
	U_{ij}   \sum_{	\alpha,\beta,
			\vec{k},\vec{k'}}
	\frac{	F[\tilde\epsilon_{\beta,\vec{k}}]+F[\tilde\epsilon_{\alpha,\vec{k'}}]-1}{\tilde\epsilon_{\beta,\vec{k}}+\tilde\epsilon_{\alpha,\vec{k'}}}
	\sum_{R_i}\sum_{R_l}\\
	&	
	\big(\overline{
		e^{i\vec{k}.(\vec{R_i}+\vec{r}_i)} \psi_{\alpha,\vec{k}}^{\mathrm{s}_i}e^{i\vec{k'}.(\vec{R_j}+\vec{r}_j)} \psi_{\beta,\vec{k'}}^{\mathrm{s}_j}
		}+\\
  &\qquad\overline{\eta 	e^{i\vec{k'}.(\vec{R_i}+\vec{r}_i)} \psi_{\beta,\vec{k'}}^{\mathrm{s}_i}e^{i\vec{k}.(\vec{R_j}+\vec{r}_j)} \psi_{\alpha,\vec{k}}^{\mathrm{s}_j}
		}\big) \times\\
	&\big(e^{i\vec{k}.(\vec{R_l}+\vec{r}_l)} \psi_{\alpha,\vec{k}}^{\mathrm{s}_l}e^{i\vec{k'}.(\vec{R_m}+\vec{r}_m)} \psi_{\beta,\vec{k'}}^{\mathrm{s}_m}
	\\\qquad&+\eta 	e^{i\vec{k'}.(\vec{R_l}+\vec{r}_l)} \psi_{\beta,\vec{k'}}^{\mathrm{s}_l}e^{i\vec{k}.(\vec{R_m}+\vec{r}_m)} \psi_{\alpha,\vec{k}}^{\mathrm{s}_m}\big)	.
\end{split}
\end{equation}
Note that in Eq.~(\ref{Mkspacemodemnt}), $R_{j}=R_{i}+R_n$ [$R_{m}=R_{l}+R_{n'}$], 
\begin{equation}\label{Mkspacemodemnt1}
\begin{split}
&M_{\mathrm{s_is_js_ls_m}}=\frac{1}{N^3}
	U_{ij}^{\eta}   \sum_{\alpha,\beta,	\vec{k},\vec{k'}}
	\frac{	F[\tilde\epsilon_{\beta,\vec{k}}]+F[\tilde\epsilon_{\alpha,\vec{k'}}]-1}{\tilde\epsilon_{\beta,\vec{k}}+\tilde\epsilon_{\alpha,\vec{k'}}}
	\times\\
	&
	\sum_{R_i} e^{-i\vec{k}.\vec{R}_i} e^{-i\vec{k'}.\vec{R}_i}\sum_{R_l}e^{i\vec{k}.\vec{R}_l}e^{i\vec{k'}.\vec{R}_{l}}\times\\
	&
	\overline{\big(
		e^{i\vec{k}.(\vec{r}_i)} \psi_{\alpha,\vec{k}}^{\mathrm{s}_i}e^{i\vec{k'}.(\vec{R_n}+\vec{r}_j)} \psi_{\beta,\vec{k'}}^{\mathrm{s}_j}
		+\eta 	e^{i\vec{k'}.(\vec{r}_i)} \psi_{\beta,\vec{k'}}^{\mathrm{s}_i}e^{i\vec{k}.(\vec{R_n}+\vec{r}_j)} \psi_{\alpha,\vec{k}}^{\mathrm{s}_j}
		\big)}
		\times\\
	&	
		\big(e^{i\vec{k}.(\vec{r}_l)} \psi_{\alpha,\vec{k}}^{\mathrm{s}_l}e^{i\vec{k'}.(\vec{R_{n'}}+\vec{r}_m)} \psi_{\beta,\vec{k'}}^{\mathrm{s}_m}
	+\eta 	e^{i\vec{k'}.(\vec{r}_l)} \psi_{\beta,\vec{k'}}^{\mathrm{s}_l}e^{i\vec{k}.(\vec{R}_{n'}+\vec{r}_m)} \psi_{\alpha,\vec{k}}^{\mathrm{s}_m}\big).
\end{split}
\end{equation}
and by summation over $R_i$ and $R_l$ we get
\begin{equation}\label{Mkspacemodemnt2}
\begin{split}
&    	M_{\mathrm{s_is_js_ls_m}}=\\
 & \frac{1}{N}
	U_{ij}^{\eta}   \sum_{\begin{matrix}
			\alpha,\beta, \vec{k}
	\end{matrix}}
	\frac{	F[\tilde\epsilon_{\beta,\vec{k}}]+F[\tilde\epsilon_{\alpha,-\vec{k}}]-1}{\tilde\epsilon_{\beta,\vec{k}}+\tilde\epsilon_{\alpha,-\vec{k}}}\times\\
 &%	\sum_{R_n}
	\overline{\big(
		\psi_{\alpha,\vec{k}}^{\mathrm{s}_i}e^{i\vec{k}.\vec{\Delta r}_{ij}} \psi_{\beta,-\vec{k}}^{\mathrm{s}_j}
		+\eta 	 \psi_{\beta,-\vec{k}}^{\mathrm{s}_i}e^{-i\vec{k}.\vec{\Delta r}_{ij}} \psi_{\alpha,\vec{k}}^{\mathrm{s}_j}
		\big)}\times
\\
&	\big(
	\psi_{\alpha,\vec{k}}^{\mathrm{s}_l}e^{i\vec{k}.\vec{\Delta r}_{lm}} \psi_{\beta,-\vec{k}}^{\mathrm{s}_m}
	+\eta 	\psi_{\beta,-\vec{k}}^{\mathrm{s}_l}e^{-i\vec{k}.\vec{\Delta r}_{lm}} \psi_{\alpha,\vec{k}}^{\mathrm{s}_m}
	\big)	
	,
\end{split}
\end{equation}
Where, $\Delta r_{ij}=\vec R_i+\vec r_i-\vec R_j-\vec r_j=\vec r_i-\vec R_n-\vec r_j$, and $\Delta \vec r_{lm}=\vec R_l+\vec r_l-\vec R_m-\vec r_m=\vec r_l-\vec R_{n'}-\vec r_m$.
 We can rewrite this equation as matrix multiplication,
\begin{equation}\label{Mmatrixproducty}
	M=\mathcal{U}^{\eta} \mathcal{B}_{\eta} \mathcal{F} \mathcal{B}_{\eta}^\dagger.
\end{equation}
The matrices in  Eq.~(\ref{Mmatrixproducty}), using Kronecker delta notation, $\delta$ are given by
\begin{equation}\label{Mmatrixproducty1}
	[\mathcal{U}^{\eta}]_{L,L'}=\delta_{L,L'} U_L,
 \end{equation}
and
 \begin{equation}
	{[\mathcal{B}_{\eta}]}_{L,E}=
%	\sum_{R_n}
\overline{	\big(
		\psi_{\alpha,\vec{k}}^{\mathrm{s}_i}e^{i\vec{k}.\vec{\Delta r}_{ij}} \psi_{\beta,-\vec{k}}^{\mathrm{s}_j}
		+\eta 	 \psi_{\beta,-\vec{k}}^{\mathrm{s}_i}e^{-i\vec{k}.\vec{\Delta r}_{ij}} \psi_{\alpha,\vec{k}}^{\mathrm{s}_j}
		\big)}, 
  \end{equation}
  and 
   \begin{equation}
	\label{Mmatrixproducty3}
	{[\mathcal{F}]}_{E,E'}=
	\frac{1}{N}\delta_{E,E'} \frac{	F[\tilde\epsilon_{\beta,\vec{k}}]+F[\tilde\epsilon_{\alpha,-\vec{k}}]-1}{\tilde\epsilon_{\beta,\vec{k}}+\tilde\epsilon_{\alpha,-\vec{k}}},
\end{equation}
where $U_{L}\equiv U_{ij}$, and we represent pair of energy  indices and vertices using $E\equiv{\alpha,\vec{k};\beta,-\vec{k}}$ and  $L\equiv{s_i;s_j}$, respectively. 
We can also rewrite the linearized gap equation in the energy basis,
\begin{equation}\label{orderparameterbasis}
	\Delta_{\alpha\beta}(\vec{k})=\sum_{i,j} \Delta^{\eta}_{ij}
		\big( \psi^{\mathrm{s}_i}_{\alpha,\vec{k}}  
	e^{i\vec{k}.\vec{\Delta r}_{ij}} \psi^{\mathrm{s}_j}_{\beta,-\vec{k}} +\eta
	 \psi^{\mathrm{s}_j}_{\alpha,\vec{k}}  
	 	e^{-i\vec{k}.\vec{\Delta r}_{ij}}
 \psi^{\mathrm{s}_i}_{\beta,-\vec{k}} 	\big).
\end{equation}
Equivalently we can write previous equation compactly as $\Delta_E= {[\mathcal{B}_{\eta}^{\dagger}]}_{E,L} \Delta_L$.

The matrix production in Eq.~(\ref{Mmatrixproducty}) is dense matrix multiplication.
Considering all of the possible energy pairs results in huge matrices. However, we can consider only couples of states in $E$, that both of them have energy within a specified window, $\omega_d$ around the chemical potential similar to standard gap equation in translationally symmetric systems.
This is because superconducting instability in the usual system is the instability of the Fermi surface.
 
 \subsection{Fulde–Ferrell–Larkin–Ovchinnikov states}

The eigenvectors of Eq.~\ref{Mmatrixproducty} consist of primary pairing instabilities and their corresponding Fulde–Ferrell–Larkin–Ovchinnikov (FFLO) states. Generally, FFLO states refer to pairing with nonzero total momentum for Cooper pairs.

In conventional crystalline superconductors with a large magnetic field, the Fermi pockets for electrons and holes with opposite spins are separated, making the FFLO state favorable. However, in some systems, finite momentum pairing can occur even without an external magnetic field. For instance, intra-valley pairing in graphene \cite{PhysRevResearch.2.043155} can lead to such states. This type of pairing breaks the original translational symmetry of the system and is considered an FFLO state \cite{PhysRevResearch.2.043155}. Furthermore, such states have also been referred to as pair-density wave states \cite{PhysRevB.94.104508}.

\begin{figure}
    \centering
    \centering\includegraphics[width=1\linewidth]{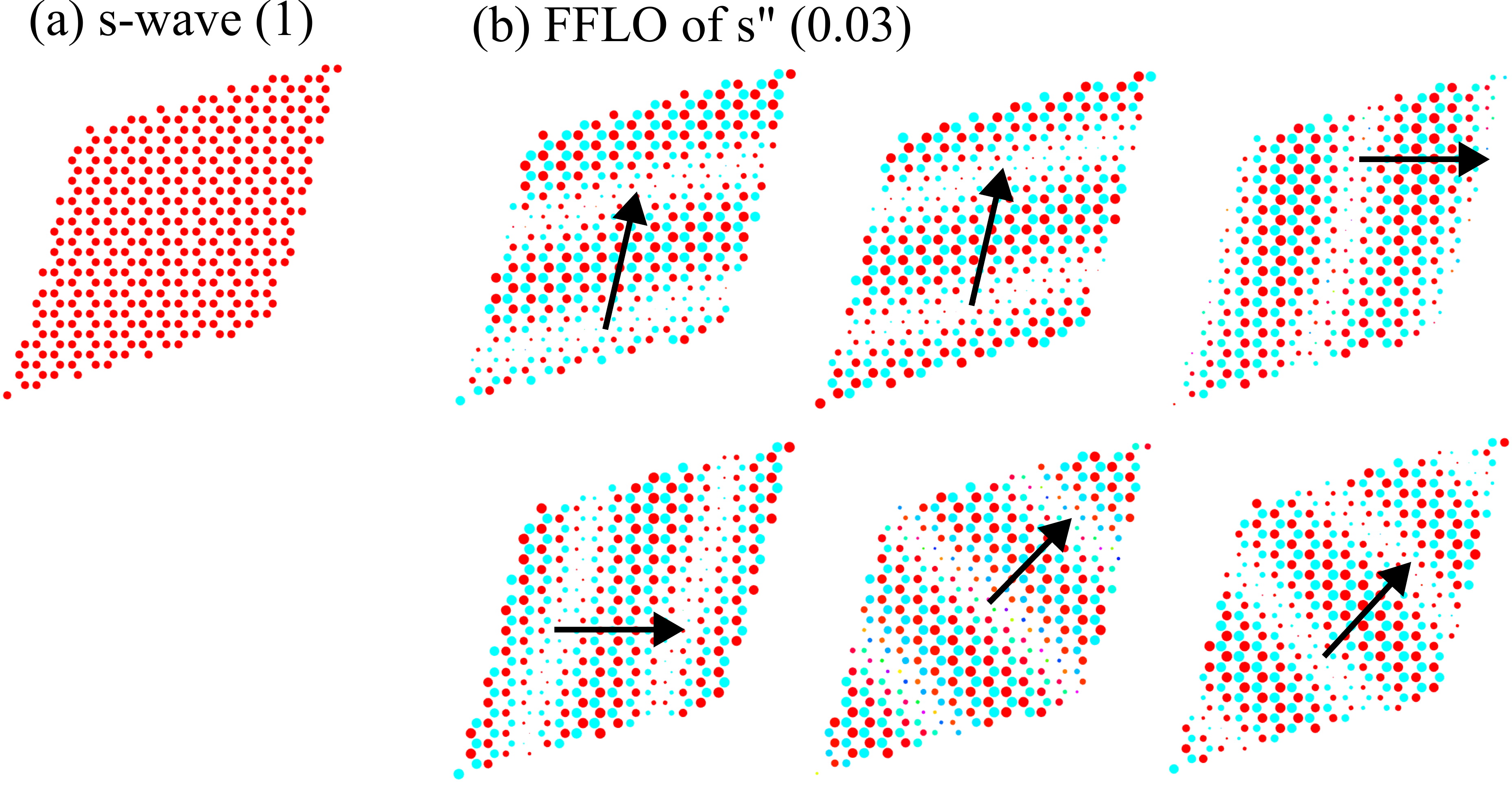}
    \caption{The (a) dominant s-wave and (b) subleading FFLO s"-wave pairing for monolayer graphene with onsite attraction and $\mu = -0.5$ eV. The s"-wave pairing is an instability with the same amplitude on both sublattices but with opposite signs. All FFLO solutions in (b) are degenerate, and the ratio of their pairing eigenvalue to that of the s-wave pairing is provided in parentheses (0.03). The arrows in (b) show the direction of spatial modulation.}
    \label{REFB-FFLO}
\end{figure}

In real space, the FFLO state implies that the pairing exhibits spatial modulation on the length scale of the lattice constant, which is related to the nonzero total momentum of Cooper pairs in the FFLO superconductor \cite{PhysRevB.94.104508}. Indeed, we observe that primary pairing states and their corresponding FFLO counterparts typically appear in the eigenvectors of the pairing matrix. For example, in Fig.~\ref{REFB-FFLO}, we plot the pairing eigenvector in monolayer graphene with only onsite attraction. We find that the most favorable pairing is uniform s-wave instability. Still, there are subleading pairing instabilities with negligible pairing eigenvalues [means they are not favorable], which locally resemble s"-wave (opposite sign for two sublattices) but modulate along certain directions (shown by arrows). Several FFLO solutions are degenerate due to the rotational symmetry of the monolayer graphene. 
Note that we can represent these FFLO states using $\Delta(r) = \cos{(i \vec{Q}_n \cdot( \vec{r} + \vec{r}_0))} \Delta_{\sigma}$, where $\Delta_{\sigma} = \sigma \Delta_0$ is the primary pairing [$\sigma=\pm1$ for the opposite sublattices]. The length of the arrows in Fig.~\ref{REFB-FFLO} is proportional to $1/|\vec{Q}_n|$, and the two solutions for each $\vec{Q}_n$ correspond to the constant shift freedom $\vec{r}_0$. In the main manuscript, we noted that these FFLO states are not important due to their small pairing eigenvalues.

It is important to note that the nature of the leading FFLO states depends on the supercell size, especially when it is small. For instance, we found that Kekulé order becomes dominant near zero energy for the next nearest neighbor spin triplet attraction when the underlying supercell respects Kekulé ordering (as shown in Fig.~\ref{fig:fig6}).

\subsection{Degenerate solution}

Sometimes multiple degenerate solutions are obtained from the linearized gap equation. 
In the phase transition, these solutions have similar free energy, when their pairing amplitude is infinitesimal.
However, by decreasing temperature and obtaining a finite value for pairing potential, a higher order of pairing potential becomes important in free energy. 
Usually, in these cases, the system brings a combination of those solutions to decrease free energy. 
If degeneracy happens between nodal pairing potential, the system picks a pairing potential that makes the superconductivity gaped. 
The new combination often breaks time-reversal symmetry and induces a nontrivial topological phase for the superconductor.
We can expand free energy up to the fourth order of pairing potential, 
\begin{equation}
\begin{split}
    	&\mathcal{F}\approx 
	\sum_{\alpha}\epsilon_{\alpha,e} F[\epsilon_{\alpha,e}]+
	\sum_{\alpha} \epsilon_{\alpha,e} F[\epsilon_{\alpha,e}]\\
 &	-\sum_{ij,\eta=\pm1}{|\Delta^{\eta}_{ij}|^2}\big/{U_{ij}}+\sum_{\alpha}\epsilon_{\alpha},
\end{split}
\end{equation}
where, by using Eqs.~(\ref{perturbationtheory})-(\ref{eigenstateperturbation1})
\begin{equation}\label{freeEnergy}
\begin{split}
    &\mathcal{F}\approx\sum_{\alpha}\frac{F[\tilde\epsilon_{\alpha}]-	F[-\tilde\epsilon_{\beta}]}{(\tilde\epsilon_{\alpha})-(-\tilde\epsilon_{\beta})}
	\Big|\sum_{\eta=\pm}\sum_{i,j} \Delta^{\eta}_{ij}	\big(	\psi^i_{\alpha}\psi^j_{\beta}+\eta \psi^j_{\alpha}\psi^i_{\beta}		\big)\Big|^2\\
 &	-\sum_{ij,\eta=\pm1}{|\Delta^{\eta}_{ij}|^2}\big/{U_{ij}}+\Delta\mathcal{F}+\text{Constant}
 \end{split}
\end{equation}
Note that to derive Eq.~(\ref{pairingpotentialsubsitut1}), we first neglect terms beyond the second order and minimize Eq.~(\ref{freeEnergy}) concerning pairing potential, $\Delta^{\eta}_{ij}$.

Now consider a general combination of degenerate pairing potential
\begin{equation}\label{combinating}
	\Delta=\sum_{i_1=1,\dots,n_d} \psi_{i_1} \Delta^{(i_1)}_{\alpha\beta},\qquad \text{where}\qquad  \sum_{i_1=1,\dots,n_d} |\psi_{i_1}|^2 =1,
\end{equation}
$n_d$, nd is the number of degenerate solutions, and  $(i_1)$ indicates different degenerate pairing potentials.
We assume that pairing is given in energy indices, which can be found through Eq.~(\ref{orderparameterbasis}).
Different combinations of pairing potential give the same free energy up to the second perturbation order.  The fourth-order correction in free energy is provided by
\begin{equation}\label{pairing4}
	\Delta\mathcal{F}=\sum_{{i_1,i_2,i_3,i_4}} \psi_{i_1}\psi_{i_2}^*\psi_{i_3}\psi_{i_4}^* \Delta\mathcal{F}_{i_1i_2i_3i_4},
\end{equation}
where
\begin{equation}\label{pairing5}
\begin{split}
 &   	\Delta\mathcal{F}_{i_1i_2i_3i_4}=\\
 &\sum_{\alpha\beta\eta\gamma}F[\tilde\epsilon_{\alpha}]
	\frac{ \Delta^{(i_1)}_{\alpha\beta}(\Delta^{(i_2)})^{*}_{\gamma\beta}\Delta^{(i_3)}_{\gamma\eta}(\Delta^{(i_4)})^*_{\alpha\eta}}{\big(\tilde\epsilon_{\alpha}+\tilde\epsilon_{\beta}\big)     \big(\tilde\epsilon_{\alpha}-\tilde\epsilon_{\gamma}\big)  \big(\tilde\epsilon_{\alpha}+\tilde\epsilon_{\eta}\big)}\\
 &	-\sum_{\alpha\beta\eta}F[\tilde\epsilon_{\alpha}]
	\frac{ \Delta_{\alpha\beta}^{(i_1)}(\Delta^{(i_2)})^{*}_{\alpha\beta}\Delta^{(i_3)}_{\alpha\eta}(\Delta^{(i_4)})^*_{\alpha\eta}}{\big(\tilde\epsilon_{\alpha}+\tilde\epsilon_{\beta}\big)^2     \big(\tilde\epsilon_{\alpha}+\tilde\epsilon_{\eta}\big)}
	\\
&	-\sum_{\alpha\beta\eta\gamma}F[-\tilde\epsilon_{\alpha}]
	\frac{ (\Delta_{\beta\alpha}^{(i_2)})^*\Delta^{(i_1)}_{\beta\gamma}(\Delta^{(i_4)})^*_{\eta\gamma}\Delta^{(i_3)}_{\eta\alpha}}{\big(\tilde\epsilon_{\alpha}+\tilde\epsilon_{\beta}\big)     \big(\tilde\epsilon_{\alpha}-\tilde\epsilon_{\gamma}\big)  \big(\tilde\epsilon_{\alpha}+\tilde\epsilon_{\eta}\big)}\\
 &	+\sum_{\alpha\beta\eta}F[-\tilde\epsilon_{\alpha}]
	\frac{ (\Delta^{(i_2)}_{\beta\alpha})^{*}\Delta^{(i_1)}_{\beta\alpha}(\Delta_{\eta\alpha}^{(i_4)})^*\Delta^{(i_3)}_{\eta\alpha}}{\big(\tilde\epsilon_{\alpha}+\tilde\epsilon_{\beta}\big)^2     \big(\tilde\epsilon_{\alpha}+\tilde\epsilon_{\eta}\big)}.
\end{split}
\end{equation}
We can find the superconducting instabilities channel with the lowest free energy by probing $\Delta \mathcal{F}$ within possible combinations of degenerate pairing potentials. Indeed we confirm that degenerate pairings always go to chiral pairing.

\clearpage
\newpage
 
\section{Supplementary Figures and Discussion}
In this section, we provide further figures and discussion to support the main paper.

\subsection{Phase diagram for different attraction channels}
\subsubsection{Trial pairing eigenvector}

\begin{figure*}[th!]
\includegraphics[width=0.7\linewidth]{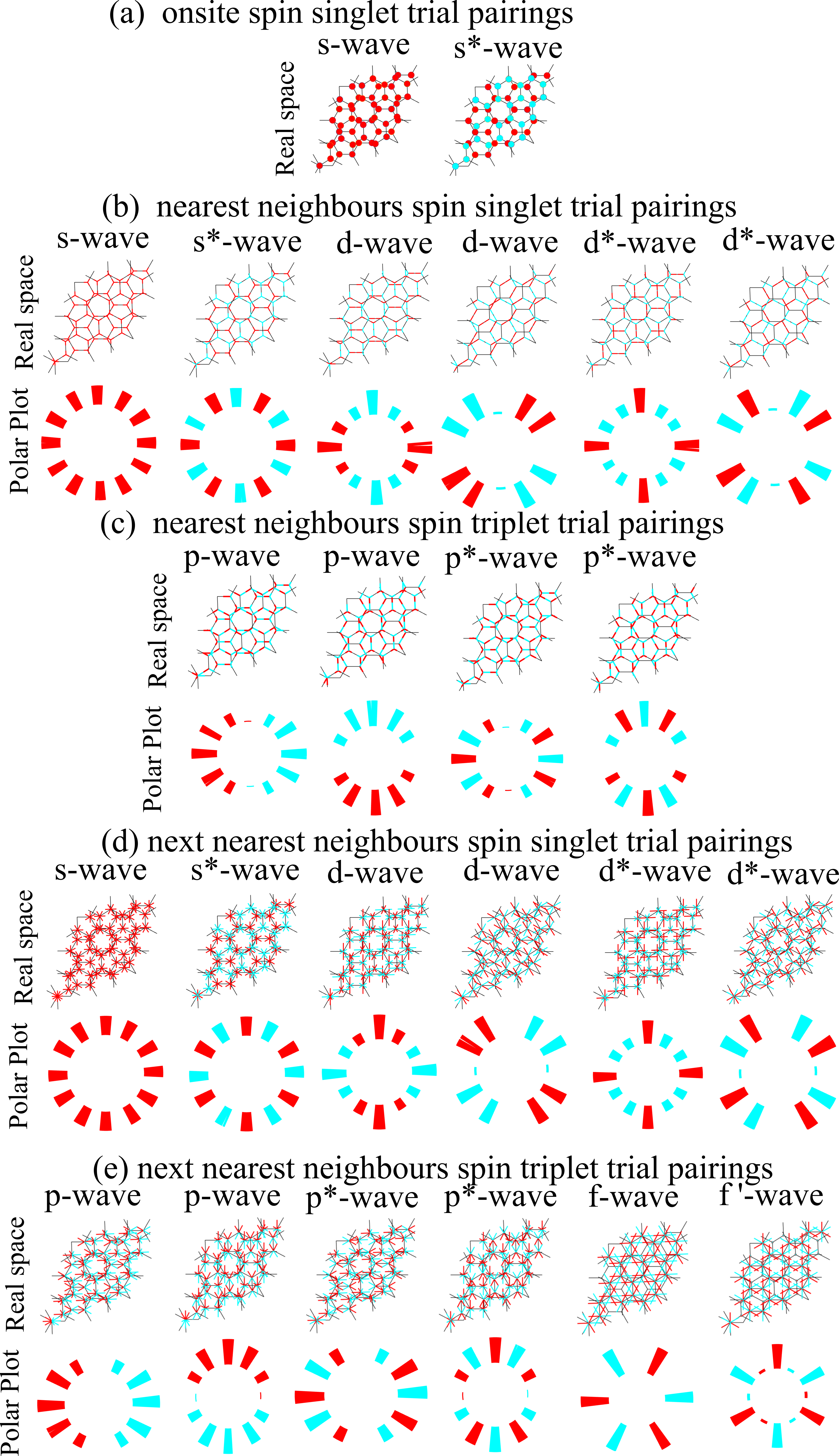}
\caption{Real space and polar plot of the distribution of trial pairing potential for $\tau_D(2)=2/1$. We plot \(\Delta_{ij}\)  by drawing a line starting from vertex \(i\) and extending along a vector connecting \(i\) to \(j\). The length of this line indicates the relative amplitude, and the colors represent the sign of the pairing (blue for negative and red for positive). 
For polar plots below some panels, we determine the polar angle that connects \(i\) to \(j\) and then draw a line in that direction, extending from \(r=r_0\) to \(r=r'>r_0\), where \(r'-r_0\) captures the relative amplitude. Similarly, the colors indicate the sign of the pairing. We also randomly rotate these lines within a small angle to avoid overlap.}
\label{fig:fig7}
\end{figure*}

The trial pairing eigenvectors are employed to determine the characteristics of the dominant pairings.  We list several trial pairings in Fig.~\ref{fig:fig7}.

For further clarification, let's denote trial eigenvectors by $\Delta^t_{\alpha}$, where $\alpha$ labels different trial pairing eigenvectors [for instance s-wave, s*-wave]. Note that $\Delta^t_{\alpha}$ are vectors whose components represent the relative pairing amplitude and phase.  The trial pairing eigenvectors can be obtained when the two layers are decoupled and aperiodicity is absent. Additionally, one can easily construct them by considering symmetrical functions. For instance,  uniform pairing with the same (or opposite) sign on both layers can serve as good trial pairing eigenvectors for onsite attraction ($[\Delta^t_{\text{s-wave}}]_{i} \propto 1$, and $[\Delta^t{\text{s*–wave}}]_{i} \propto l$, where the subscript $i$ denotes the component of the onsite pairing on vertex $i$, and $l = \pm$ indexes the layers [see Fig.~\ref{fig:fig7}(a)]). 
Similarly, for the nearest-neighbor attraction, trial eigenvectors can be constructed using $[\Delta^t_{\text{p}_x}]_{ij} \propto \cos \theta_{ij}$, $[\Delta^t_{\text{p}^*_x}]_{ij} \propto l \cos \theta_{ij}$, $[\Delta^t_{\text{p}_y }]_{ij} \propto \sin \theta_{ij}$, and $[\Delta^t_{\text{p}^*_y }]_{ij} \propto l \sin \theta_{ij}$, where $\theta_{ij}$ are the polar angle for the link connecting the nearest-neighbor sites $i$ and $j$ [see Fig.~\ref{fig:fig7}(c)]. Here, $[\Delta^t_{\alpha}]_{ij}$ specifies the vector component corresponding to the pairing between vertices $i$ and $j$.  

Now, when the two layers are coupled and aperiodicity may exist, the linearized gap equation gives the dominant and subdominant pairing instabilities, which we denote by $\Delta_j$ [For instance, we can take $j = 1$ for the dominant pairing with the highest pairing eigenvalue, and $j = 2, 3, \dots$ for the subleading pairings with smaller pairing eigenvalues.] 

Note that $\Delta^t_{\alpha}$ and $\Delta_j$ are normalized vectors, as the linearized gap equation cannot determine the pairing amplitude. The pairing overlap amplitude (POA) is given by their inner product $|\Delta^*_j \cdot\Delta^t_{\alpha}|^2$ and determines how much of $\Delta_j$ can be described by $\Delta^t_{\alpha}$. Note that, in Fig.3 (g-i) we report the summation of POAs for trial function with the same symmetry. For instance, for p-wave, p*-wave, d-wave, and d*-wave symmetry,  there are two trial pairing eigenvectors [e.g. p$_x$, and p$_y$].

\FloatBarrier

\subsubsection{Phase diagrams}

\begin{figure*}
\includegraphics[width=1\linewidth]{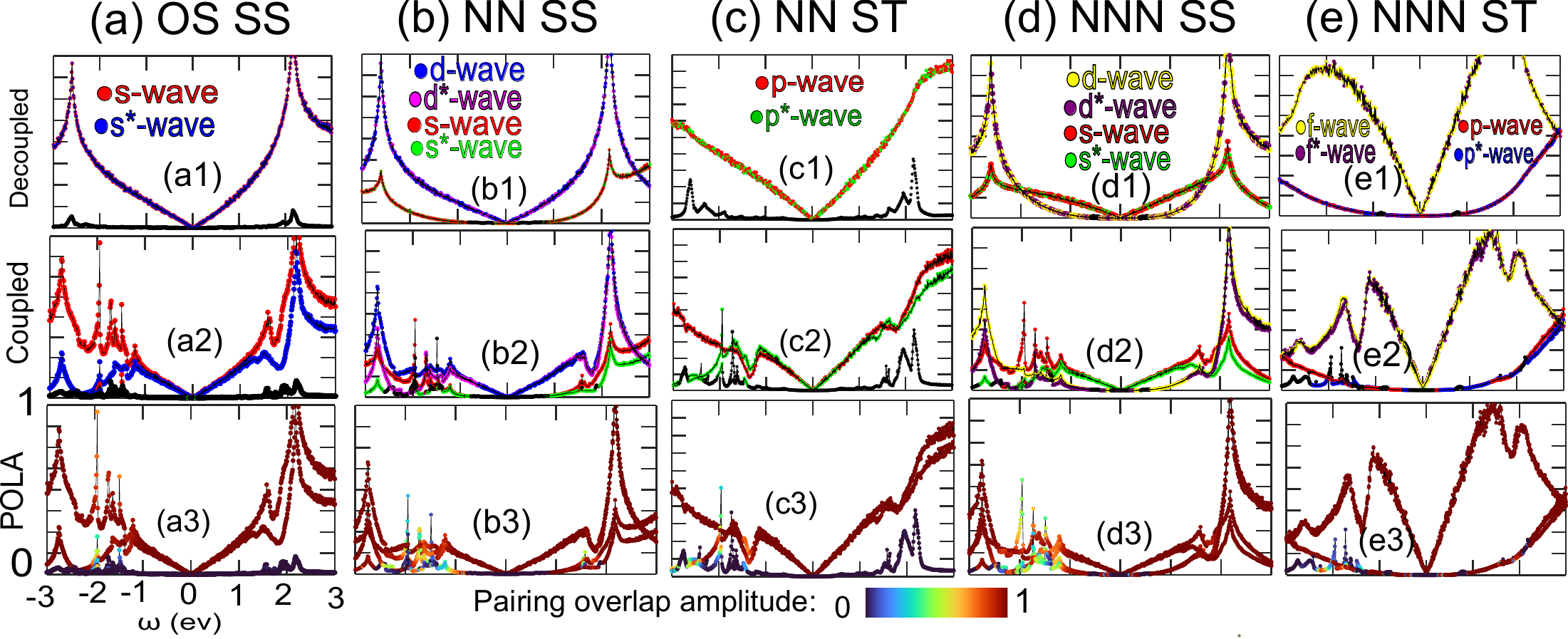}
\caption{ (a1-e1, a2-e2) The largest eigenvalue of $M$ is plotted versus chemical potential $\mu=\omega$. The larger eigenvalue indicates a higher possibility of the superconducting instability towards the corresponding eigenvector.
To understand, the symmetry of a given eigenvector, we calculate overlap of it with already known trial pairing potential. We appoint the characteristics of the given eigenvector as the trial pairing potential with the largest overlap amplitude (POA). 
The colors show the characteristics of corresponding pairing instabilities. 
The black indicates that no trial pairing potential can describe the given pairing eigenvector.
The calculation in (a1-e1) [(a2-e2)] is given for decoupled (coupled) graphene layers.
In (a3-e3) similar to (a2-e2), we plot the largest eigenvalue of $M$, but colors show the value of the largest POA. Dark brown indicates perfect overlap with a high symmetric pairing potential, while colors going toward blue indicate less POA, and dark blue (and black) indicate no overlap. 
In the inset, we zoom in on the energy interval $\omega\in[-2.5,-1] \text{eV}$, where the vertical axis shows POA and color shows the corresponding symmetry of the most probable superconducting instability.
The POA is almost one in any range of energy except the aperiodical regime of the energy spectrum. 
When POA is approaching one (zero), the pairing potential distribution is even (strongly fluctuating).
The attraction channels for each column is written on the top and are (a) onsite spin singlet [OS SS], (b) nearest neighbor spin singlet [NN SS], (c) nearest neighbor spin triplet [NN ST], (d) next nearest neighbor spin singlet [NNN SS], and (e) next nearest neighbor spin triplet [NNN ST], respectively.
} 
\label{fig:fig3SM}
\end{figure*}

In Fig.~\ref{fig:fig3SM}, we summarize all calculations in a single figure for different channels of pairings. 
In Fig.~\ref{fig:fig3SM}(a), the attraction is onsite and leads to s-wave on the broad range of the chemical potential. 

In Fig.~\ref{fig:fig3SM}(b), we consider the nearest neighbor spin singlet attraction channel, where mostly d-wave is dominant. Note, however, that in some QCvHS, d*-wave or s-wave can be selected. Here, due to underlying symmetry, d-wave and d*-wave have a double degeneracy ($d_{x^2-y^2}$ with $d_{xy}$ or $d^*_{x^2-y^2}$ with $d^*_{xy}$).

\begin{figure*}[t]
\includegraphics[width=0.65\linewidth]{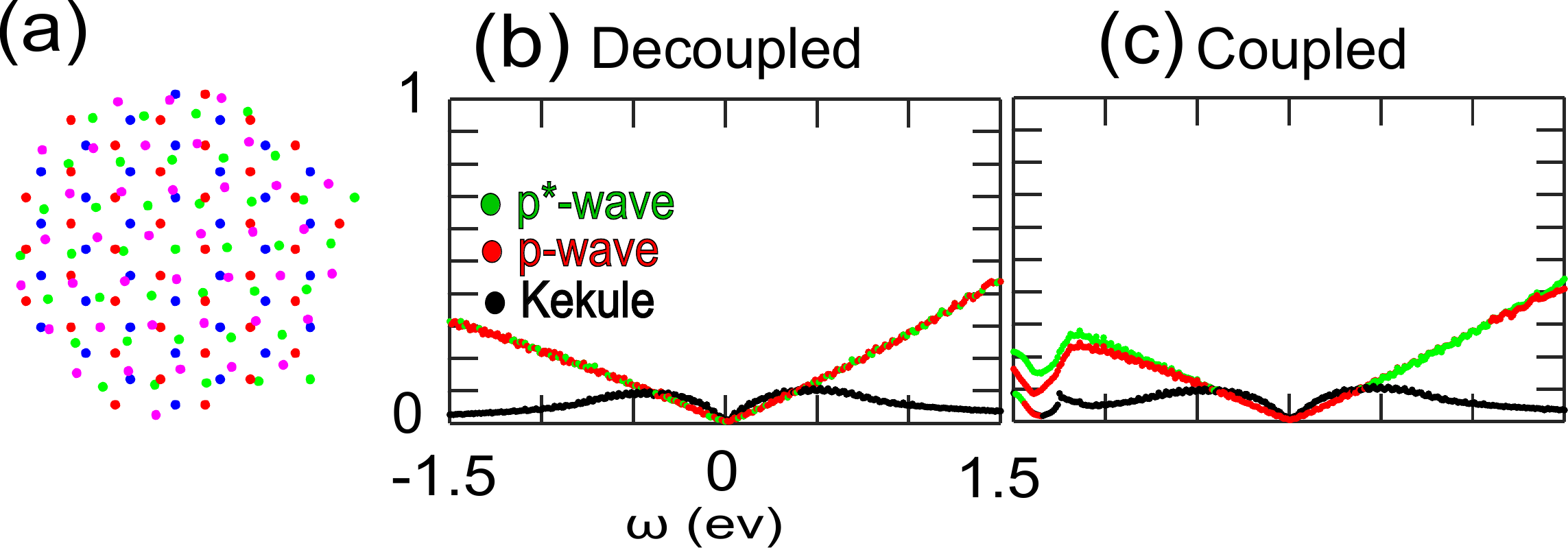}
\caption{Dominant pairing eigenvalue for nearest neighbor spin triplet attraction with respecting Kekule unit cell ($\theta=32.2042$) shows four-fold degenerate (2 layers and 2 valleys) Kekule superconductivity (Black curve) around zero energy. (a) approximant supercell. (b) coupled and (c) decoupled phase diagram.}
\label{fig:fig6}
\end{figure*}

In Fig.~\ref{fig:fig3SM}(c), we consider the nearest neighbor spin triplet attraction channel, where mostly p-wave or p*-wave is dominant. 
Due to underlying symmetry, p-wave and p*-wave have a double degeneracy ($p_x$ with $p_y$ or $p^*_x$ with $p^*_y$).
Note that the approximant that we use here and in the main paper does not respect the Kekule unit cell. However, one can obtain the same angle rotated quasicrystal by starting with the Kekule unit cell and rotating it. In Fig.~\ref{fig:fig6}, we found that near zero energy in the nearest neighbor spin triplet attraction, and for supercell respecting Kekule ordering, pairing with Kekule ordering appears, which has a 2*2 degeneracy (2 layers and 2 valleys).  This pairing is intera-Dirac cone, breaking the original translational symmetry and having finite total momentum.  Near zero energy, the Fermi surface becomes circular, allowing pairing within one valley with constant total momentum and consequently enhancing the appearance of Kekule superconductivity.

In Fig.~\ref{fig:fig3SM}(d), we consider the next nearest neighbor spin singlet attraction channel, where mostly s-wave is dominant. Note, however, that near the main VHS, d-wave or d*-wave is most favorable. 

In Fig.~\ref{fig:fig3SM}(e), we consider the next nearest neighbor spin triplet attraction channel, where mostly f-wave is dominant. Note that d-wave (or d*-wave) and p-wave (or p*-wave) are double degenerate. The f-wave pairing is nearly degenerate due to slight deviation (due to approximant), and we expect dual degeneracy in the quasicrystal limit due to $D_{6d}$ symmetry. Note that chiral pairing near zero energy is not valuable or important as two mono-layer graphenes are already gapped without chiral pairing. However, in larger doping where f-wave pairing may give nodal superconductivity for the mono-layer case, $f \pm if$ can become more favorable and remove nodes in TBGQC.

In general, for attraction that contains all neighbors with different distances and a mixture of triplet and singlet, their relative amplitude of attraction, and the resulting eigenvalue can determine what is the dominant pairing symmetry. 

Near zero energy, effectively, layers are weakly coupled, and the pairing eigenvalue difference between d-wave (p-wave or s-wave) and d*-wave (p*-wave or s*-wave) is small. However, the pairing eigenvalue difference is larger in QER.

\subsection{Distribution of pairing eigenvector for different channels of attraction }
In the following, we plot \(\Delta_{ij}\) to illustrate the distribution of the pairing eigenvector in real space. We do this by drawing a line starting from vertex \(i\) and extending along a vector connecting \(i\) to \(j\). The length of this line indicates the relative amplitude, and the colors represent the sign of the pairing (blue for negative and red for positive). 

To better understand the symmetry of the pairing, we also use polar plots. For this, we determine the polar angle that connects \(i\) to \(j\) and then draw a line in that direction, extending from \(r=r_0\) to \(r=r'>r_0\), where \(r'-r_0\) captures the relative amplitude. Similarly, the colors indicate the sign of the pairing. We also randomly rotate these lines within a small angle to avoid overlap.

In Fig.~\ref{fig:fig7}(b-e), we present both the real space and polar plots of trial pairings. Note the following characteristics:
\begin{itemize}
    \item s-wave: Exhibits a constant color in all directions.
\item  s*-wave: Alternates in color in the polar plot.
\item d-wave and d*-wave:  Opposite directions have the same color, and there are four and eight nodes in the polar plot.
\item p-wave and p*-wave: Opposite directions have the opposite color, and for p-wave two nodes are present.
\item  f-wave: Opposite directions have the opposite color and six nodes exist.
\end{itemize}
These plots help visualize various pairing states.
Furthermore, as pairing is uniform through the lattice structure in the decoupled limit, we can see that in polar and real space plots there is no fluctuation while we can see later that in QER there is. 

\begin{figure*}[t]
\includegraphics[width=0.7\linewidth]{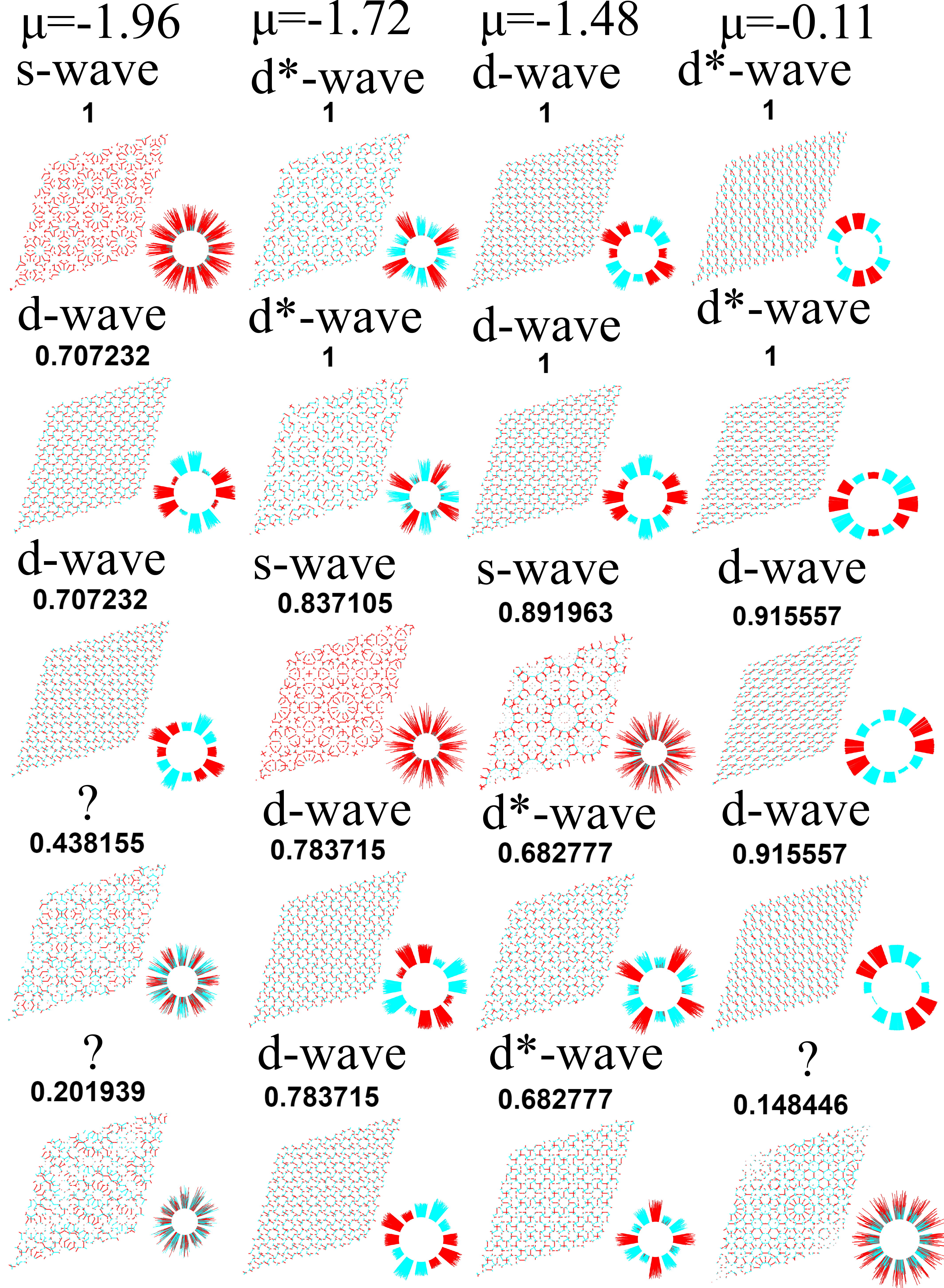}
\caption{Real space and polar plot of the distribution of pairing for nearest neighbor spin-singlet pairing attraction. We plot \(\Delta_{ij}\)  by drawing a line starting from vertex \(i\) and extending along a vector connecting \(i\) to \(j\). The length of this line indicates the relative amplitude, and the colors represent the sign of the pairing (blue for negative and red for positive). 
For polar plots below some panels, we determine the polar angle that connects \(i\) to \(j\) and then draw a line in that direction, extending from \(r=r_0\) to \(r=r'>r_0\), where \(r'-r_0\) captures the relative amplitude. Similarly, the colors indicate the sign of the pairing. We also randomly rotate these lines within a small angle to avoid overlap.}
\label{fig:NNsinglet}
\end{figure*}

\begin{figure*}[t]
\includegraphics[width=1\linewidth]{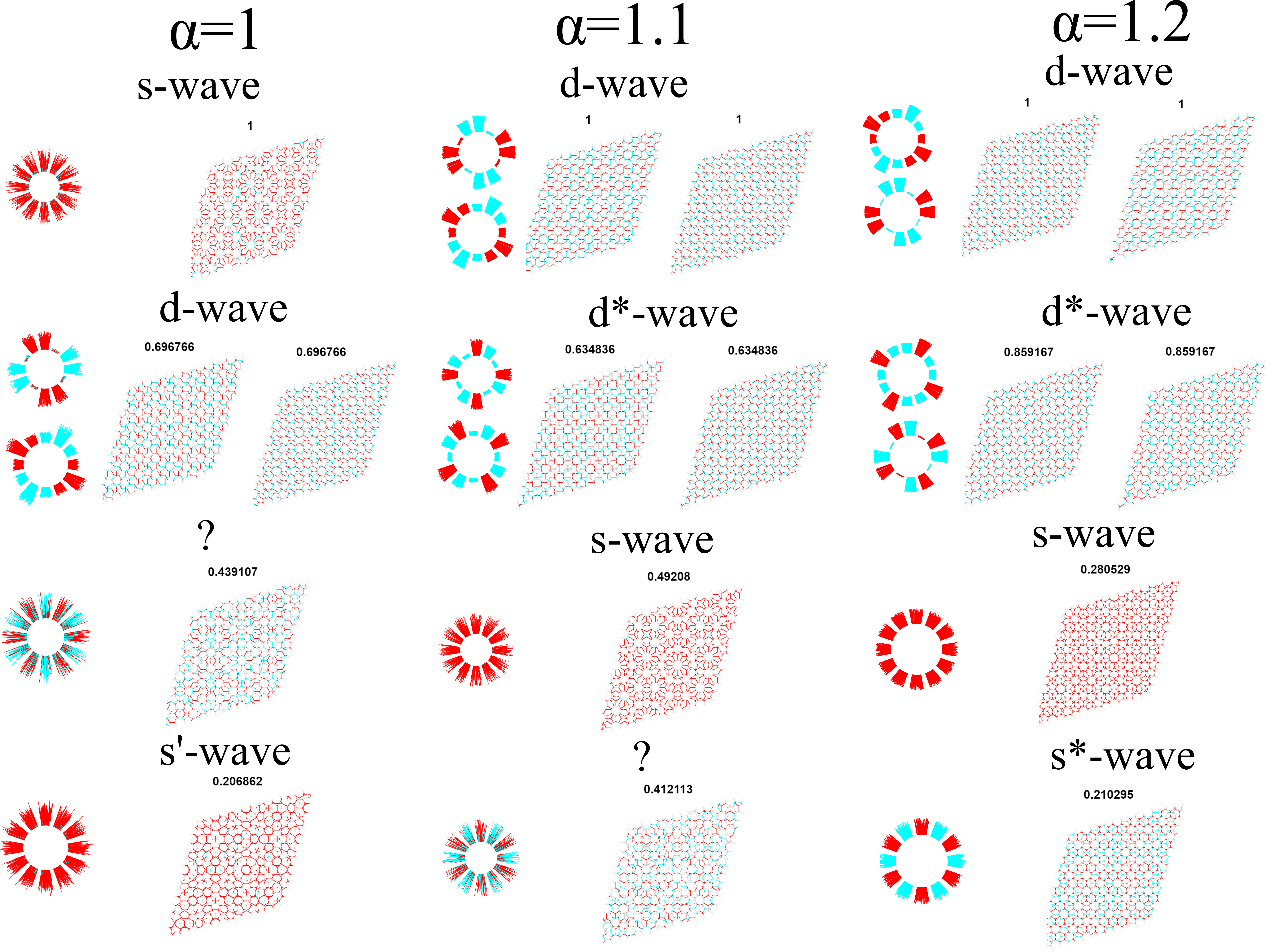}
\caption{  Evolution of dominant pairings versus layer decoupling $h_z\rightarrow \alpha h_z$, where $\alpha=1$ is TBGQC and $\alpha\gg 1$ is two decoupled graphene layers. We choose $\mu=-1.96$. As we can see in large $\alpha$,  the pairing eigenvalue difference of $d$ and $d*$ is suppressing.  Furthermore, aperiodic s-wave pairing is sensitive to quasiperiodic strength which decreases by increasing layer distance. We plot \(\Delta_{ij}\)  by drawing a line starting from vertex \(i\) and extending along a vector connecting \(i\) to \(j\). The length of this line indicates the relative amplitude, and the colors represent the sign of the pairing (blue for negative and red for positive). 
For polar plots below some panels, we determine the polar angle that connects \(i\) to \(j\) and then draw a line in that direction, extending from \(r=r_0\) to \(r=r'>r_0\), where \(r'-r_0\) captures the relative amplitude. Similarly, the colors indicate the sign of the pairing. We also randomly rotate these lines within a small angle to avoid overlap. }
\label{fig:figdwave}
\end{figure*}

In Fig.~\ref{fig:NNsinglet}, we plot the dominant pairing in different chemical potentials. First, in QER, the distribution of pairing is position dependent, and a large fluctuation in the polar plot is seen. While in PER ($\mu=-0.11$) it is even. Note that at $\mu=-1.96$ s-wave type of pairing was established, which by deviating chemical potential quickly suppressed its eigenvalue. The number in each plot shows the relative eigenvalue compared to the largest eigenvalue.  In Fig.~\ref{fig:figdwave}, we plot the evolution of dominate paring by increasing layer distance, where quasiperiodicity is decreasing. We can see that the aperiodic s-wave pairing is very sensitive to the quasi-periodicity strength. Furthermore, the difference between d-wave and d*-wave is suppressed by increasing layer distance.  Note that since layer distance is important for stabilizing aperiodic s-wave pairing, we expect that TBGQC may exhibit aperiodic superconductivity or even phase transition under the application of vertical pressure.

\begin{figure*}[t]
\includegraphics[width=0.7\linewidth]{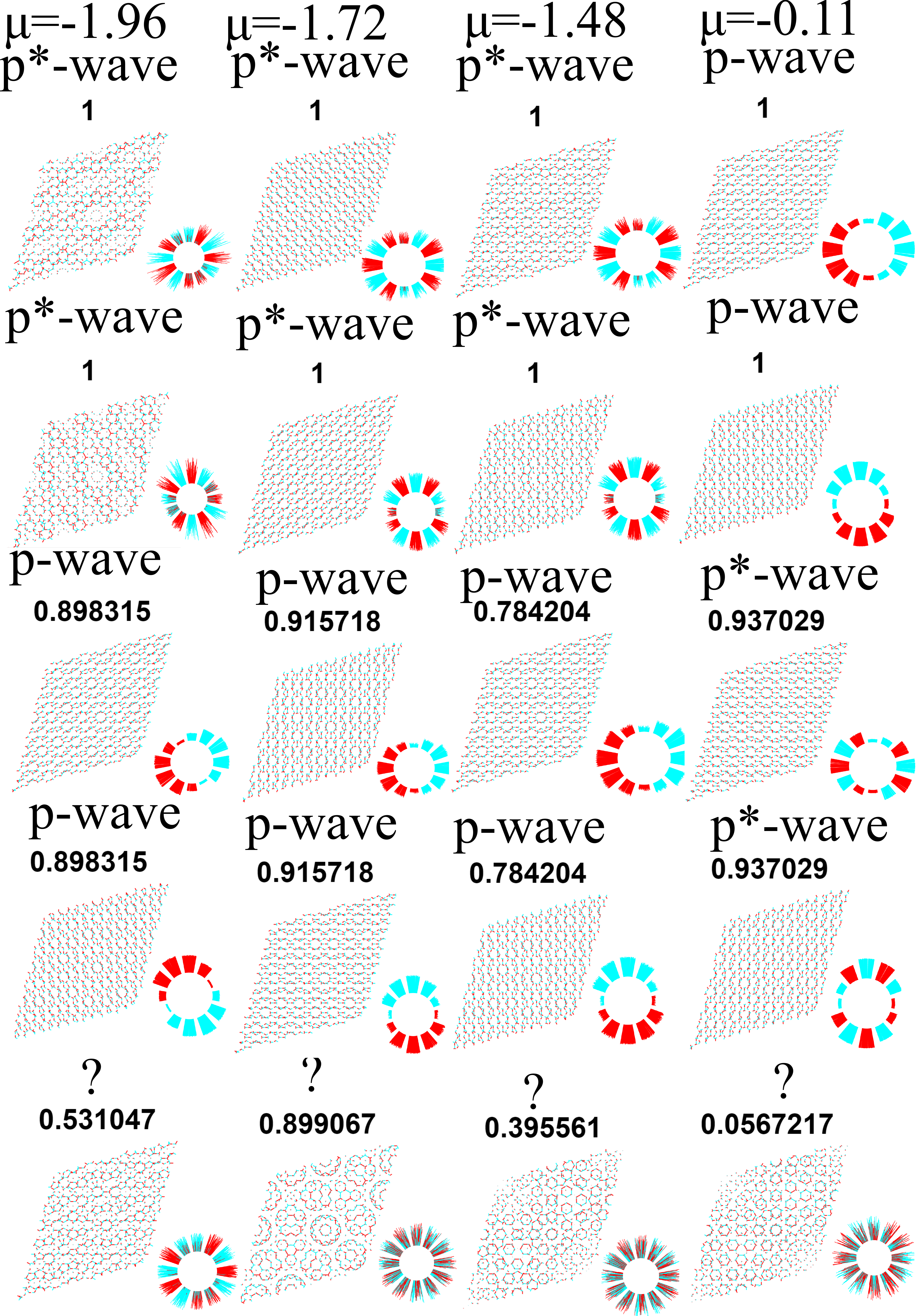}
\caption{Real space and polar plot of the distribution of pairing for nearest neighbor spin-triplet pairing attraction. We plot \(\Delta_{ij}\)  by drawing a line starting from vertex \(i\) and extending along a vector connecting \(i\) to \(j\). The length of this line indicates the relative amplitude, and the colors represent the sign of the pairing (blue for negative and red for positive). 
For polar plots below some panels, we determine the polar angle that connects \(i\) to \(j\) and then draw a line in that direction, extending from \(r=r_0\) to \(r=r'>r_0\), where \(r'-r_0\) captures the relative amplitude. Similarly, the colors indicate the sign of the pairing. We also randomly rotate these lines within a small angle to avoid overlap.}
\label{fig:NNtriplet}
\end{figure*}

In Fig.~\ref{fig:NNtriplet} we plot dominant pairing for nearest neighbor spin-triplet pairing attraction. we can see strong aperiodicity for p*-wave  in $\mu=-1.96$. While generally, week aperiodicity exists for p-wave or p*-wave in other QER, in PER ($\mu=-0.11$) it is completely absent. 

\begin{figure*}[h]
\includegraphics[width=0.7\linewidth]{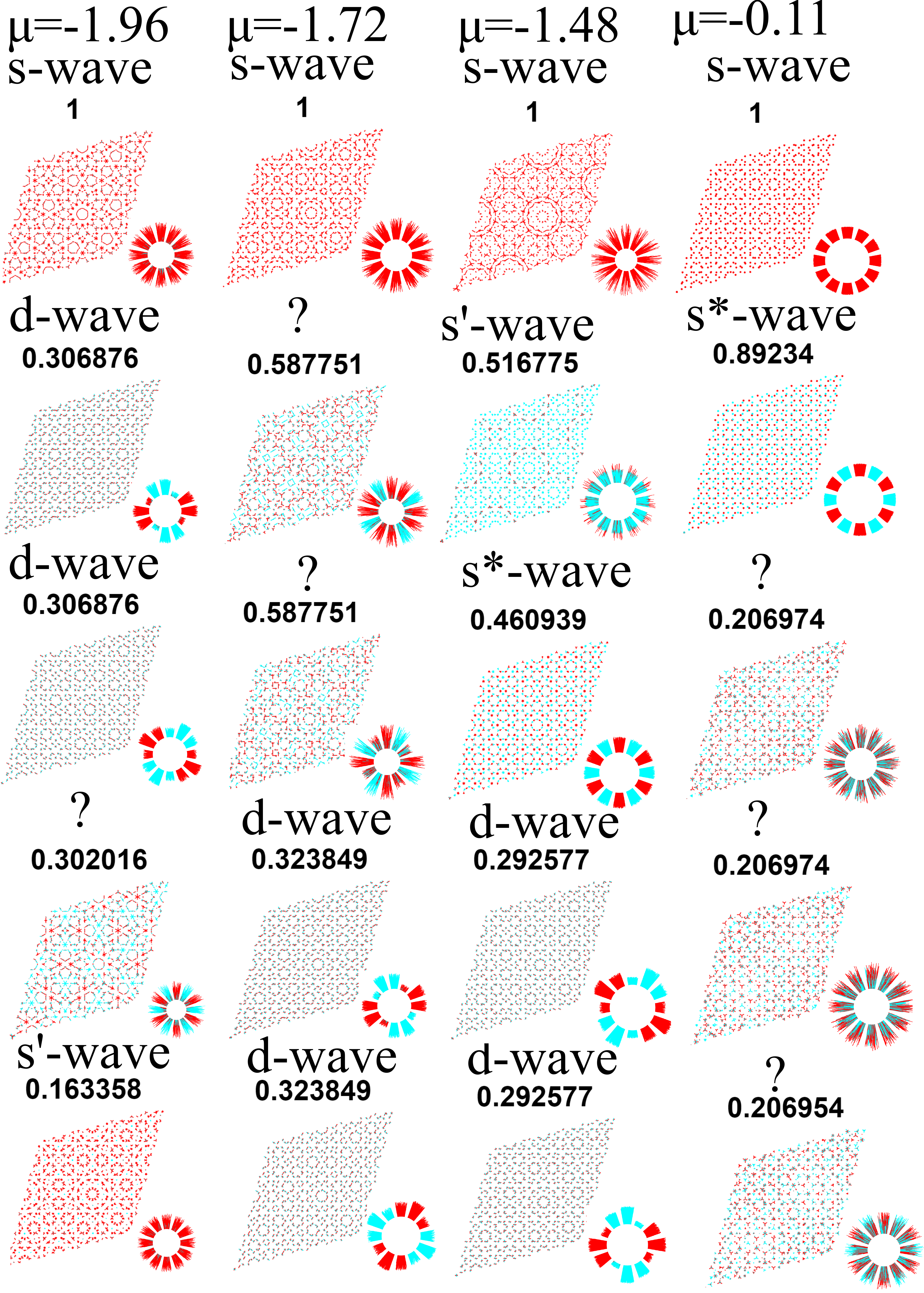}
\caption{Real space and polar plot of the distribution of pairing for next nearest neighbor spin-singlet pairing attraction. We plot \(\Delta_{ij}\)  by drawing a line starting from vertex \(i\) and extending along a vector connecting \(i\) to \(j\). The length of this line indicates the relative amplitude, and the colors represent the sign of the pairing (blue for negative and red for positive). 
For polar plots below some panels, we determine the polar angle that connects \(i\) to \(j\) and then draw a line in that direction, extending from \(r=r_0\) to \(r=r'>r_0\), where \(r'-r_0\) captures the relative amplitude. Similarly, the colors indicate the sign of the pairing. We also randomly rotate these lines within a small angle to avoid overlap.}
\label{fig:NNNSinglet}
\end{figure*}

In Fig.~\ref{fig:NNNSinglet} we plot dominant pairing for the next nearest neighbor spin-singlet pairing attraction. We can see strong aperiodicity for s-wave in QER, while in PER constant s-wave pairing is dominated. 

\begin{figure*}[t]
\includegraphics[width=0.7\linewidth]{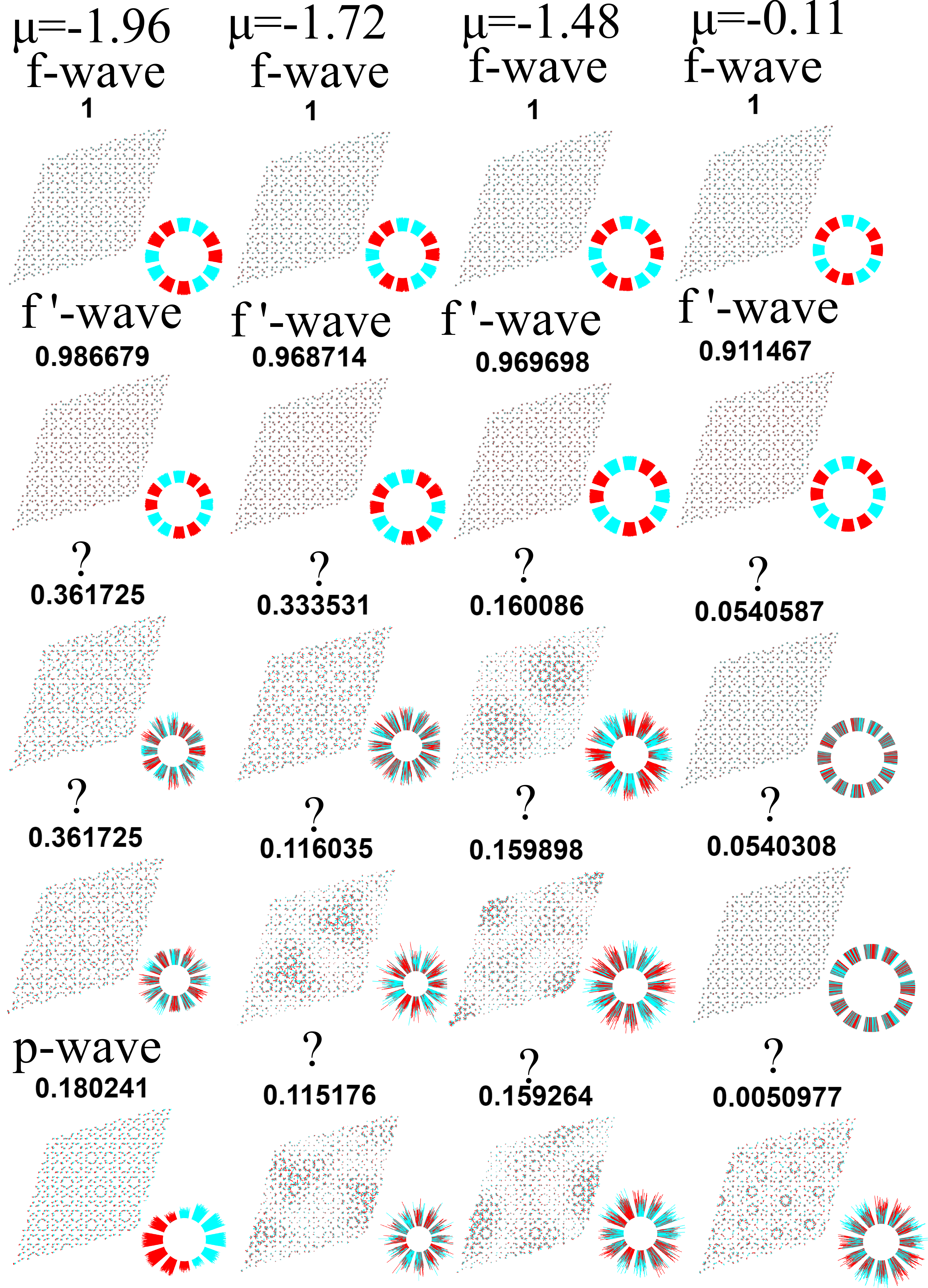}
\caption{Real space and polar plot of the distribution of pairing for next nearest neighbor spin-triplet pairing attraction. We plot \(\Delta_{ij}\)  by drawing a line starting from vertex \(i\) and extending along a vector connecting \(i\) to \(j\). The length of this line indicates the relative amplitude, and the colors represent the sign of the pairing (blue for negative and red for positive). 
For polar plots below some panels, we determine the polar angle that connects \(i\) to \(j\) and then draw a line in that direction, extending from \(r=r_0\) to \(r=r'>r_0\), where \(r'-r_0\) captures the relative amplitude. Similarly, the colors indicate the sign of the pairing. We also randomly rotate these lines within a small angle to avoid overlap.}
\label{fig:NNNTriplet}
\end{figure*}
\FloatBarrier

In Fig.~\ref{fig:NNNTriplet} we plot dominant pairing for the next nearest neighbor spin-triplet pairing attraction. we can see for the dominant f-wave pairing, no aperiodicity exists for the whole energy range. Note that the first and second f-wave pairings have slight eigenvalue differences and in the quasicrystal limit we expect they become degenerated due to underlying $D_{6d}$ symmetry. 
	
\subsection{Discussion on topological superconductivity in TBGQC}
Close to zero energy, two layers are weekly coupled and therefore the degeneracy of dominant pairing is nearly doubled (e.g. $d_{xy}$,  $d_{x^2-y^2}$, $d^*_{xy}$,  and $d^*_{x^2-y^2}$). 
 On the other hand, it is known that for monolayers, both $d$-wave and $p$-wave show sign changes on the Fermi surface and therefore lead to nodal pairing superconductivity ~\cite{Black-Schaffer_2014}.  
 However, degenerate nodal pairings are unstable and the system can lower its free energy by obtaining gapped chiral pairing (e.g. $d_{xy}+id_{x^2-y^2}$ or $d_{xy}-id_{x^2-y^2}$) instabilities from them \cite{Kallin_2016, Nandkishore2012, PhysRevLett.111.066804, Black-Schaffer_2014,PhysRevB.108.134515}. 
 This leads to the spontaneous breaking of time-reversal symmetry and the emergence of gapped topological superconductivity with nontrivial Chern number $\mathcal{C}\neq 0$.    
 Although we expect chiral pairings are always favorable for each monolayer superconductor, layers can obtain either the same chiral pairing (e.g. $d_{xy}+id_{x^2-y^2}$ for the top layer and $d_{xy}+id_{x^2-y^2}$ for the bottom layer) or the opposite chiral pairing (e.g. $d_{xy}+id_{x^2-y^2}$ for the top layer and $d_{xy}-id_{x^2-y^2}$ for the bottom layer). 
	We found that for the same chiral pairing of p-wave and d-wave (with and without $^*$), the Bott index (see Sec.~\ref{sec:Bott}), which is equivalent to the Chern number in quasicrystals \cite{loring2019guide,PhysRevB.104.144511}, is twice that of the monolayer graphene case \cite{PhysRevB.108.134515}  ($|\mathcal{C}|=2\times2$).
	Although the Bott index is vanishing for the opposite chiral pairing,  it may still lead the system to exhibit higher-dimensional topological phases similar to Kekule and f-wave pairing instabilities   \cite{Li_2022,PhysRevB.105.195149,PhysRevB.107.224511}.  
 Note that in QER, d-wave has larger pairing eigenvalue than d*-wave; therefore, unique same chiral pairing can be achieved for both layers.

\subsection{Degree of aperiodicity}
Consider a pairing matrix in the decoupling limit. We diagonalize this matrix to obtain different pairing eigenvectors \(|\Delta^0_i\rangle\) and their corresponding eigenvalues \(\zeta_i^0\). 

When we introduce layer coupling as a perturbation, we modify the pairing matrix by adding a perturbative term \(M'\) to the original matrix \(M\) [Note that this separation is also perturbatively and we cannot generally do it as it constructed as a complex multiplication of wave functions]. In perturbation theory, the perturbed eigenvalues \(\zeta_i\) and eigenvectors \(|\Delta_i\rangle\) are calculated as corrections to the unperturbed solutions.

For the first-order correction to the eigenvalues, we have:
\begin{equation}\label{eigenvaluediffernece}
    \zeta_i \approx \zeta_i^0 + \text{eig} (\langle \Delta^0_i | M' | \Delta^0_{i'} \rangle),
\end{equation}
where $i$ and $i'$ are in degenerate pairing (in layer decoupled limit we have always layer degeneracy) instabilities, and "eig" refers to eigenvalues. 
 However, despite degenerate first-order perturbation can explain the change in pairing eigenvalues, it can not explain the inhomogeneity of the pairing eigenvectors, as all degenerate \(\Delta_i\) states have uniform distribution with similar symmetry but on different layers.

To understand the effect on the eigenvectors, we have to use perturbation theory that takes all other instabilities:
\[
|\Delta_j\rangle \propto  |\Delta^0_j\rangle + \sum_{\zeta_i \ne \zeta_j} \frac{\langle \Delta^0_i | M' | \Delta^0_j \rangle}{\zeta_j^0 - \zeta_i^0} |\Delta^0_i\rangle
\]
This expression shows that the first-order correction to the eigenvector \(|\Delta^0_j\rangle\) is a sum over the contributions from all other states \(|\Delta^0_i\rangle\) weighted by the perturbation matrix element \(\langle \Delta^0_i | M' | \Delta^0_j \rangle\) and the inverse of the pairing eigenvalue difference \((\zeta_j^0 - \zeta_i^0)\). Two factors are crucial here:
\begin{enumerate}
    \item \textbf{Numerator}: The matrix element \(\langle \Delta^0_i | M' | \Delta^0_j \rangle\) indicates matrix element between different channels of pairing due to perturbed pairing matrix and is proportional to the strength of the coupling between the layers.
    \item \textbf{Denominator}: The pairing eigenvalue difference \(\zeta_j^0 - \zeta_i^0\) shows how close or far apart are the pairing instabilities in their pairing eigenvalues.
\end{enumerate}

For capturing aperiodicity effects beyond eigenvalue changes, we must consider how the perturbed eigenvectors \(|\Delta^1_j\rangle\) acquire characteristics from other states. This depends on both the strength of the perturbation or layer coupling (numerator) and the proximity of the pairing eigenvalues (denominator).

Note that non-periodical and inhomogeneous pairings are typically at \(\zeta_i^0 = 0\). Therefore, for achieving aperiodical pairing for $| \Delta_i \rangle$, the pairing eigenvalue \(\zeta_j^0\) must be small. Hence, larger \(\zeta_j^0\) requires a stronger layer coupling via \(M'\) to significantly perturb and mix these states with \(\zeta_i^0 = 0\).

In the context of the NNN-ST pairing (Fig.~\ref{fig:fig3SM}(e)), f-wave pairings are generally separated significantly from other pairing instabilities and it becomes maximized at QER (energy range where layer coupling is strong and gives larger \(\langle \Delta^0_i | M' | \Delta^0_j \rangle\)). Maximizing pairing eigenvalue in QER means that for the f-wave pairing, it is less likely to be perturbed aperiodically. We found that f-wave pairings almost don't show aperiodicity, even in QER (see Fig.~\ref{fig:NNNTriplet}). 
% Note that f-wave and f'-wave are nearly degenerated, and it is natural to expect suppression of eigenvalue due to layer coupling as we anticipated from first-order perturbation given in Eq.~\ref{eigenvaluediffernece}.

Conversely, in other attraction channels such as NN-SS (Fig.~\ref{fig:fig3SM}(b)), NNN-SS (Fig.~\ref{fig:fig3SM}(d)), pairing eigenvalues do not become maximized at QER and therefore they are more susceptible to become aperiodical pairings. 

\subsection{ Temperature and generation dependence of the phase diagram}

\begin{figure*}[t!]
	\includegraphics[width=0.7\linewidth]{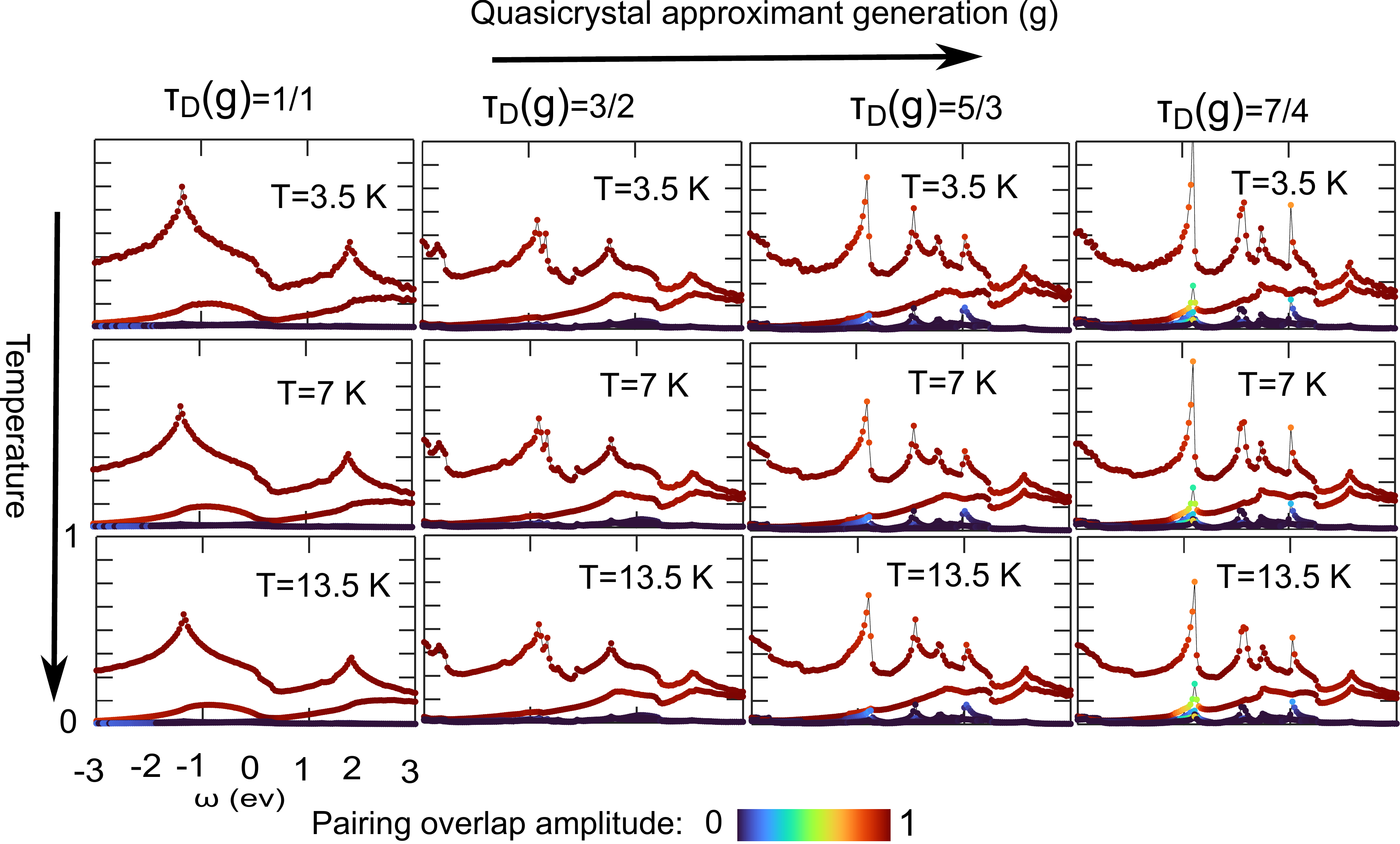}
	\caption{Pairing eigenvalue of dominant pairing potentials for spin-singlet onsite attraction channel versus temperature $T_c$ and approximant generation.  The color bar shows pairing overlap amplitude with s-wave or s*-wave. With increasing approximant generation, the system becomes more quasiperiodic. Especially near some QCvHS pairing overlap decreases and superconductivity becomes more aperiodical. Changing temperature does not change qualitatively the superconducting phase diagram.}
	\label{fig:fig5}
\end{figure*}

In Fig.~\ref{fig:fig5}, we found that increasing approximant generation enhances quasiperiodic superconductivity near QCvHS. This can be observed by noting that near QCvHS, the pairing amplitude (with s-wave and s*-wave) decreases. Increasing the temperature in Equation~\ref{Mmatrixproducty3} for the calculation does not change the phase diagram and overall fact that in QCvHS, the pairing overlap amplitude decreases.

\FloatBarrier

\subsection{Environment based classification space}
Intrinsic quasicrystals can be constructed by cut-and-projection from higher dimensions. In this method, a subset of points is projected onto a surface in a higher-dimensional space, creating the quasicrystal. These points are typically selected based on their image in a region within the perpendicular space. It is known that this perpendicular space can classify points based on their local environments and is extensively used to analyze how local aperiodic environments influence order parameters, wave functions, and more [Phys. Rev. B 102, 224201 (2020)]. 

\begin{figure*}[t!]
	\centering\includegraphics[width=0.75\linewidth]{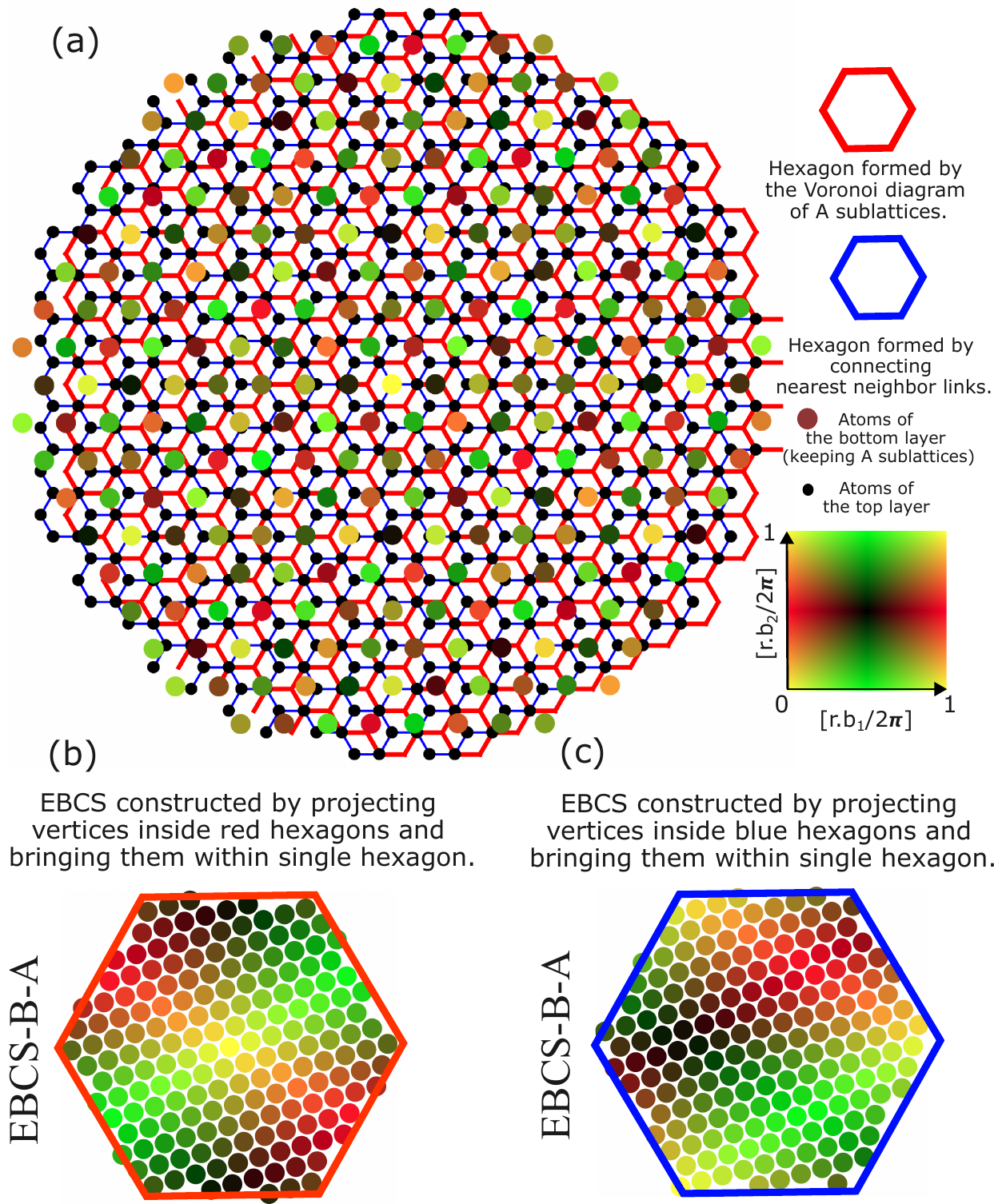}
	\caption{ Environment based classification space (EBCS) is created by projecting vertices from the bottom layer onto the hexagonal lattice of the top layer. In (a), A-sublattice vertices of the bottom layer are colored based on their relative positions within the unit cell of the top layer (see 2D color bar, where edges have identified colors). The hexagonal lattice can be chosen for instance as (b) the Voronoi diagram (red) or (c) the nearest-neighbor links (blue) of the top layer. Here, EBCS-B-A means that the EBCS is constructed for the bottom layer and A sublattice. }
	\label{fig:SMEBCS}
\end{figure*}

However, we lack direct access to this perpendicular space for classifying points by local environment in graphene quasicrystals, which are extrinsic quasicrystals. 
Different local environments can be identified by determining the relative position of vertices with a specific sublattice index (e.g., A sublattice) and layer (e.g., bottom layer) within the unit cell of the other layer (top layer) [note that in Fig.2(b) and Fig.3(e) of the main manuscript, EBCS plots are for vertices with A sublattice on the top layer]. In Fig.~\ref{fig:SMEBCS}(a), we color the A vertices of the bottom layer based on their relative positions inside the unit cell of the top layer, using $[\vec{b}_1 \cdot \vec{r} / 2\pi]$ and $[\vec{b}_2 \cdot \vec{r} / 2\pi]$ \footnote{
Suppose $\vec{r} = \alpha \vec{a}_1 + \beta \vec{a}_2$, where $\alpha=\vec{b}_1 \cdot \vec{r} / 2\pi$ and $\beta=\vec{b}_2 \cdot \vec{r} / 2\pi$ are fractional coordinates of $\vec r$ on the top layer that describes by its lattice vectors $\vec{a}_1$  and $\vec{a}_2$. The values of $[\alpha]$ and $[\beta]$ thus determine the position of $\vec{r}$ within the unit cell defined by the vectors $\vec{a}_1$ and $\vec{a}_2$.
}, where $\vec{r}$ is the position of the vertex, $\vec{b}_i$ are reciprocal lattice vectors of the top layer, and $[]$ denotes modulo 1.

We can create the EBCS by projecting vertices onto the hexagonal lattice of the other layer. The choice of hexagonal lattice remains somewhat arbitrary.  In Fig.~\ref{fig:SMEBCS}(a), the red hexagon is derived from the Voronoi diagram~\footnote{
A Voronoi diagram partitions space into regions around a set of points, where each region contains all points closest to its associated point. In the case of a triangular lattice composed solely of A sublattices (with B sublattices excluded), the Voronoi diagram generated by the A vertices results in a hexagonal lattice. } of the A sublattices in the top layer, while the blue hexagon is created by connecting nearest-neighbor links of the top layer. By projecting vertices of the bottom layer onto either the red [Fig.~\ref{fig:SMEBCS}(b)] or blue [Fig.~\ref{fig:SMEBCS}(c)] hexagons of the top layer, we obtain EBCS representations that effectively capture the environmental characteristics of the atoms. Note that both EBCS capture the same information and are connected by a constant shift of the hexagons. 

\begin{figure*}[t!]
	\centering\includegraphics[width=0.75\linewidth]{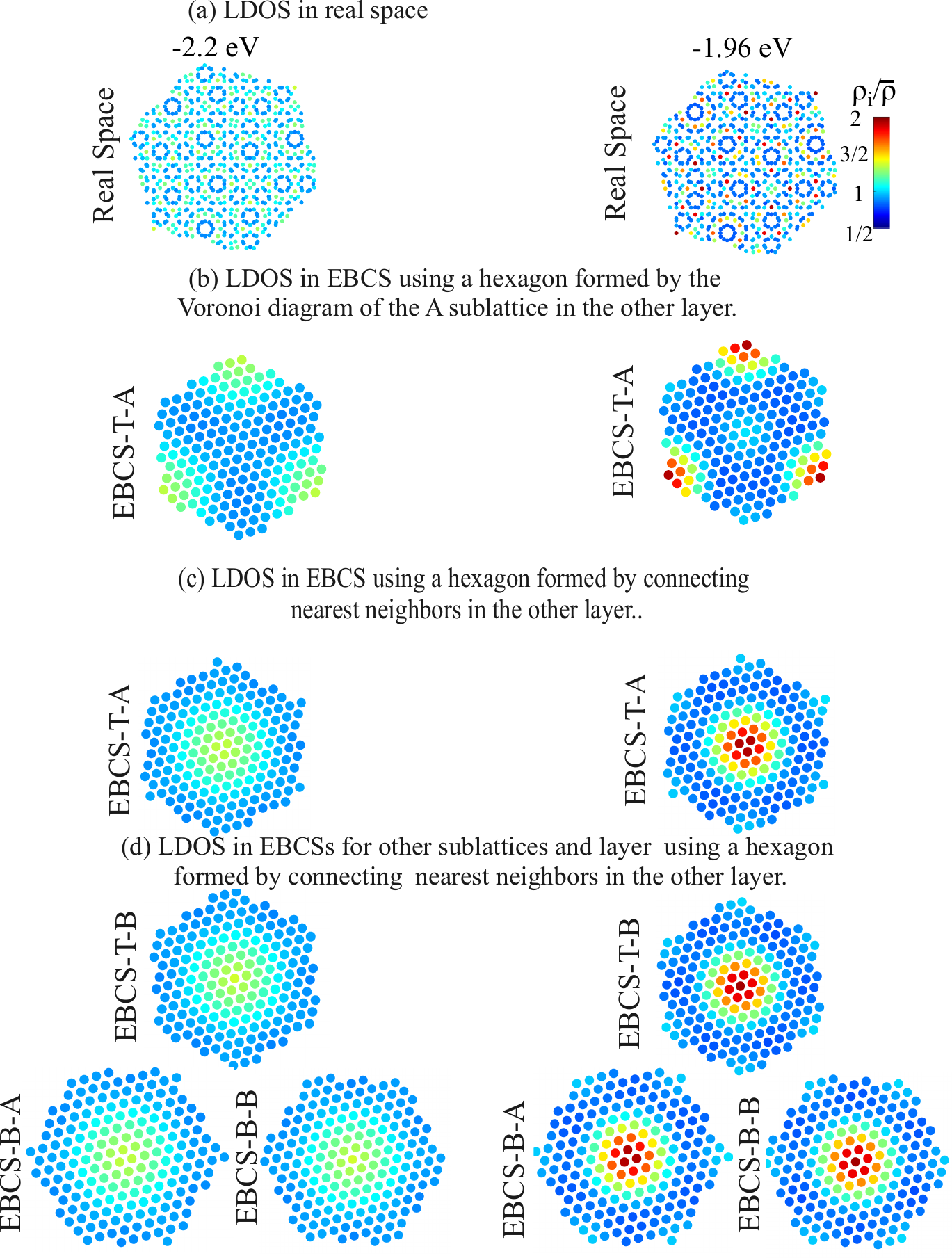}
	\caption{Local density of states (LDOS) plots. (a) The real-space plot of LDOS for two different energies (left and right column). (b) EBCS plot of (a) using hexagons created by the Voronoi diagram of A vertices in the other layer. We are showing only EBCS of A vertices of the top layer. (c) EBCS plot of (a) using hexagons formed by connecting nearest neighbors links on the other layer. (d) Same as (c), but for other sublattices and layers. We name each EBCS using the notation EBCS-X-Y, where the first letter (X) indicates whether it belongs to the bottom (X = B) or top layer (X = T), and the second letter (Y) specifies whether it belongs to the A (Y = A) or B lattice (Y = B). }
	\label{fig:SMEBCSCOMPARISON}
\end{figure*}

\begin{figure*}[t!]
	\centering\includegraphics[width=0.55\linewidth]{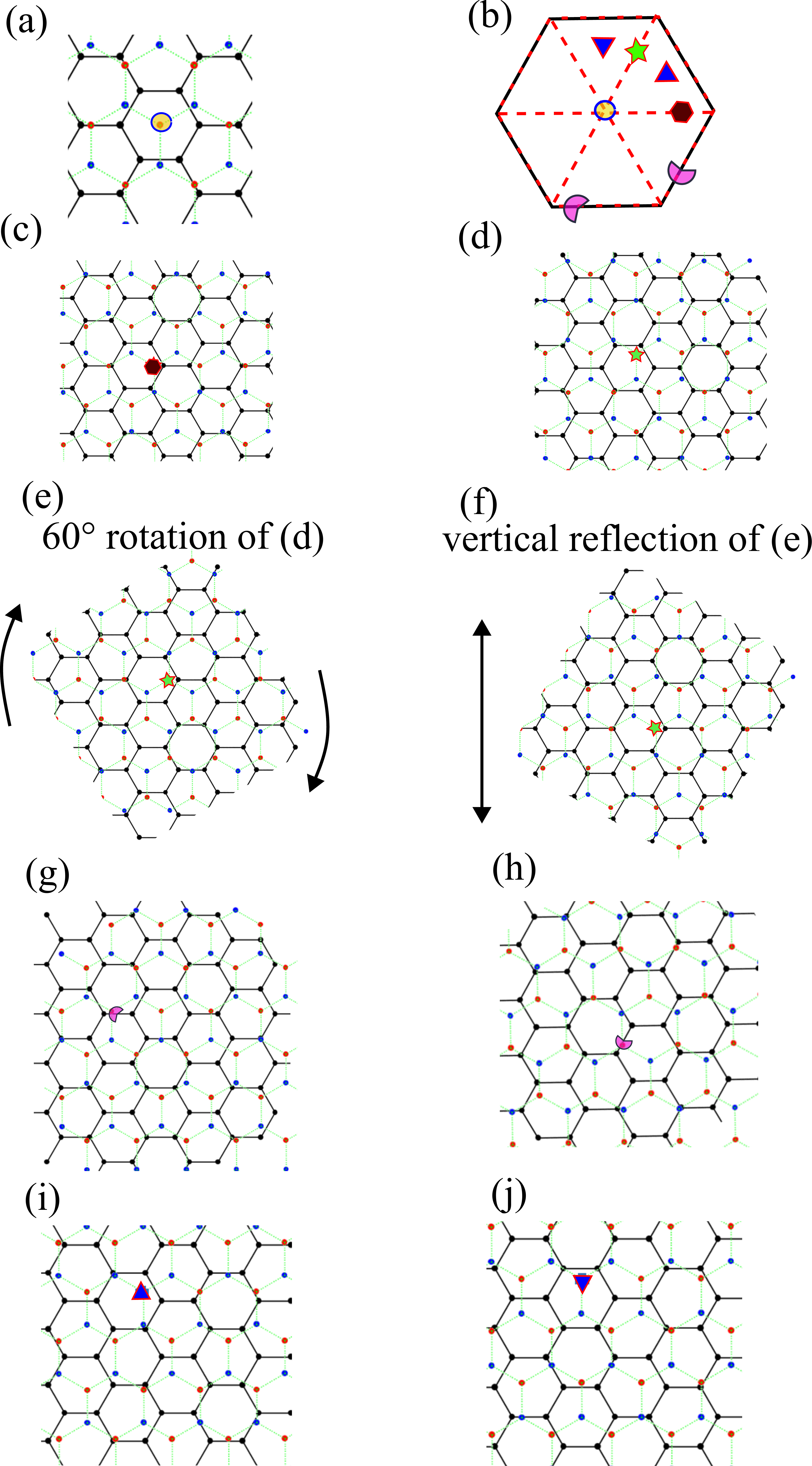}
	\caption{EBCS Symmetry. (a) The three-fold symmetric local environment is shown. (b) Seven distinct positions with different symbols are displayed in the EBCS for a B sublattice (red vertices) in the bottom layer (dotted hexagonal lattice), corresponding to the local environments shown in (a) disk symbol, (c) hexagon symbol, (d) star symbol, (g, h) crescent symbols, and (i, j) triangle symbols. The red dashed line indicates regions in the EBCS where the local environments are equivalent to the local environment of the other point in the EBCS [related by a $60^\circ$ rotation]. (c, d) Similar local environments correspond to two vertices in different positions in the EBCS [indicated in (b) by the hexagon and star]. The configuration in (d) transforms to (c) after a (e) $60^\circ$ rotation and (f) vertical reflection. (g, h) Similar local environments for two vertices in symmetric positions on the EBCS [shown in (b) as crescent shapes]. (i, j) Different local environments for two vertices in general positions in the EBCS [indicated in (b) by triangle shapes]. }
	\label{fig:SMEBCSSymmetry}
\end{figure*}

However, we found that the hexagon created by connecting nearest neighbors in the second layer better reflects the symmetry of the underlying structure. For instance, compare Fig.~\ref{fig:SMEBCSCOMPARISON} (b), and Fig.~\ref{fig:SMEBCSCOMPARISON} (c), which map LDOS in real space Fig.~\ref{fig:SMEBCSCOMPARISON} (a) to EBCS with hexagon made by A sublattice Voronoi diagram, and nearest neighbor links, respectively. In the later representation, the EBCS center corresponds to a situation, where A vertices are in the center of the other layer hexagonal lattice [Fig.~\ref{fig:SMEBCSSymmetry}(a)], while the corners indicate full eclipse with other layer vertices. 

Similarly, we construct the EBCS for the B sublattice vertices using the same hexagons formed by the other layer's nearest-neighbor links. The same procedure is applied to the A and B vertices of the other layer as well. Therefore there are four EBCS regarding two layers and two sublattices.  As shown in Fig.~\ref{fig:SMEBCSCOMPARISON}(c-d), the EBCSs for the A and B sublattices in both layers display an identical pattern.  In Fig.~\ref{fig:SMEBCSCOMPARISON}, we name each EBCS using the notation EBCS-X-Y, where the first letter (X) indicates whether it belongs to the bottom (X = B) or top layer (X = T), and the second letter (Y) specifies whether it belongs to the A (Y = A) or B sublattices (Y = B). 
In the main manuscript plots [Fig.2 (b), and Fig.3 (e)], we use the EBCS that is constructed by nearest-neighbor links.

The three-fold symmetry of the EBCS plots is related to the underlying three-fold symmetry of the system. For instance consider the central vertex [shown by a disk] of Fig.~\ref{fig:SMEBCSSymmetry}(a).
Moving the central vertex in Fig.~\ref{fig:SMEBCSSymmetry}(a) along three nearby blue vertices produces a similar environment and therefore EBCS has three-fold rotational symmetry [see following discussion].  The six-fold or twelve-fold rotational symmetries of the graphene quasicrystal change the sublattice or layer index and, therefore, cannot dictate any symmetry on each EBCS plot. However, these symmetries account for the identical patterns observed for the four EBCSs on different sublattices and layers.

Interestingly, we found that EBCS plots also have approximate six-fold symmetry [see Fig.~\ref{fig:SMEBCSCOMPARISON}(c,d)].
We analyze the existence of such symmetry, by plotting the local environment regarding different positions in EBCS [see Fig.~\ref{fig:SMEBCSSymmetry}]. 
we found that certain positions of EBCS have additional six-fold symmetry as they effectively describe the same local environment [red dashed regions in Fig.~\ref{fig:SMEBCSSymmetry}(b)]. To see it, consider two atoms with different EBCS positions [shown by a hexagon and star in Fig.~\ref{fig:SMEBCSSymmetry}(b)], but related with $60^\circ$ rotation. Their corresponding local environment are plotted in Fig.~\ref{fig:SMEBCSSymmetry}(c) and Fig.~\ref{fig:SMEBCSSymmetry}(d).
It is easy to see that, indeed the configuration in Fig.~\ref{fig:SMEBCSSymmetry}(d) can be converted to Fig.~\ref{fig:SMEBCSSymmetry}(c) by a $60^\circ$ rotation [Fig.~\ref{fig:SMEBCSSymmetry}(e)] and vertical reflection [Fig.~\ref{fig:SMEBCSSymmetry}(f)]. 
We also confirm that the local environments in Fig.~\ref{fig:SMEBCSSymmetry}(g), and Fig.~\ref{fig:SMEBCSSymmetry}(h) corresponding to crescent shape in Fig.~\ref{fig:SMEBCSSymmetry}(b) are convertible. 
Therefore, as they have a similar environment, they should produce the same LDOS, and consequently, EBCS plots display six-fold rotational symmetry for the region near red dashed lines in Fig.~\ref{fig:SMEBCSSymmetry}(b). 
Note that EBCS in other places rather than the red dashed line in Fig.~\ref{fig:SMEBCSSymmetry}(b) do not have exact six-fold symmetry. 
For instance, the local environments in Fig.~\ref{fig:SMEBCSSymmetry}(g), and Fig.~\ref{fig:SMEBCSSymmetry}(h) corresponding to triangles in Fig.~\ref{fig:SMEBCSSymmetry}(b) are not convertible. Therefore EBCSs have approximate six-fold symmetry. Additionally, the approximant size can affect the patterns of plots.

\begin{figure*}[t!]
	\includegraphics[width=0.75\linewidth]{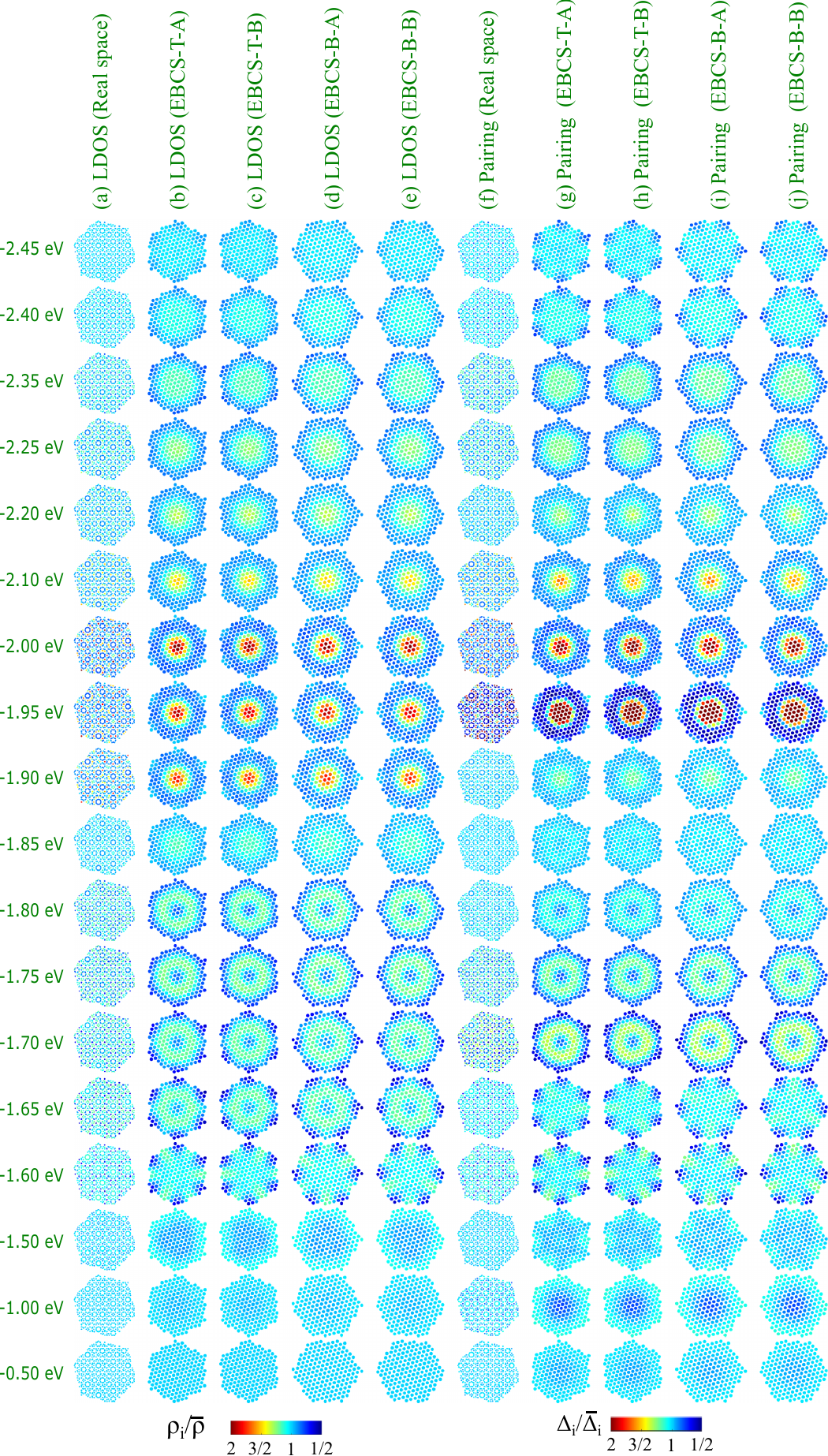}
	\caption{(a-e) LDOS in (a) real space and (b-e) EBCS. (f-j) Pairing profile for onsite effective attraction in (f) real space and (g-j) EBCS. The color bar shows the relative values of LDOS and pairing, normalized by their respective averages ($\bar{\Delta}_i=\tfrac{1}{N}\sum_i \Delta_i$, $\bar{\rho}_i=\tfrac{1}{N}\sum_i \rho_i$).}
	\label{fig:SMEBCSLDOSPAIRING}
\end{figure*}

In Fig.~\ref{fig:SMEBCSLDOSPAIRING}, we plot the LDOS [Fig.~\ref{fig:SMEBCSLDOSPAIRING}(a-e)] and pairing profile [Fig.~\ref{fig:SMEBCSLDOSPAIRING}(f-j)] for various energies in both real space [Fig.~\ref{fig:SMEBCSLDOSPAIRING}(a,f)] and EBCS [Fig.~\ref{fig:SMEBCSLDOSPAIRING}(b,c,d,e,g,h,i,j)].

\FloatBarrier

\subsection{Discussion on the phase diagram}

We note that the general features of the phase diagram of TBGQC are mainly dictated by the superconductivity of the monolayer graphene. The first row of Fig.~\ref{fig:fig3SM} shows the superconducting phase diagram of monolayer graphene [no interlayer coupling]. Therefore, it is valuable to understand phase diagrams for monolayer graphene first. Below, we first discuss the case of monolayer graphene and then return to the two-layer superconductivity. 

In crystalline systems, the attraction can be decomposed into symmetric functions, allowing for the calculation of the effective attraction for each channel. Consider an interaction described by:
\begin{equation}\label{Attraction}
    H=\sum_{ij}U_{ij}\big(c^\dagger_{i\uparrow}c^\dagger_{j\downarrow} \pm c^\dagger_{j\uparrow}c^\dagger_{i\downarrow}\big)\big(c_{j\downarrow}c_{i\uparrow} \pm c_{i\downarrow}c_{j\uparrow}\big),
\end{equation}
In monolayer graphene,  we can rewrite this Hamiltonian in momentum space using
\begin{equation}\label{transformation}
    c_{i\uparrow}=\sum_{\mathbf k}\psi^{i*}_{\mathbf k \alpha}c_{\mathbf k\alpha\uparrow}, \quad
    c_{i\downarrow}=\sum_{\mathbf k}\psi^{i*}_{\mathbf k \alpha}c_{\mathbf k\alpha\downarrow}, \quad
    c^\dagger_{i\uparrow}=\sum_{\mathbf k}\psi^i_{\mathbf k \alpha}c^\dagger_{\mathbf k\alpha\uparrow}, \quad
    c^\dagger_{i\downarrow}=\sum_{\mathbf k}\psi^i_{\mathbf k \alpha}c^\dagger_{\mathbf k\alpha\downarrow}.
\end{equation}
Here, $\alpha$ denotes energy bands. Since the superconductivity is a Fermi surface instability, we only consider the energy branch that intersects with the chemical potential and omits the energy index $\alpha$ in the following. The $\psi^i_{\mathbf k \alpha}$ is given by:
\begin{equation}
    \psi^i_{\mathbf k \alpha}=\tfrac{1}{\sqrt{N}} \sum_k \mathfrak{U}_{\sigma_i \mathbf k} e^{i \mathbf{k} \cdot \mathbf{r}_i},
\end{equation}
where $\sigma_i$ is the sublattice index of $i$. By assuming an effective attraction for certain links, such as onsite or nearest-neighbors, we can write $U_{ij}=-\delta(|\mathbf{r}_i-\mathbf{r}_j|-\mathcal{D}) U$ ($U > 0$) [e.g. $\mathcal{D}=0$ for onsite attraction].

Rewriting the Hamiltonian in momentum space and retaining only the terms relevant for zero-momentum Cooper pairs, we find:
\begin{align} \label{Attractionmomentum}
    H = -\tfrac{U}{N^2} \sum_i \sum_{n} \sum_{k,k'} &\big(
    \mathfrak{U}^*_{\sigma_i \mathbf k} e^{-i \mathbf{k} \cdot \mathbf{r}_i}
    \mathfrak{U}^*_{\sigma_j -\mathbf k} e^{i \mathbf{k} \cdot \mathbf{r}_j}
    \pm \mathfrak{U}^*_{\sigma_j \mathbf k} e^{-i \mathbf{k} \cdot \mathbf{r}_j}
    \mathfrak{U}^*_{\sigma_i -\mathbf k} e^{i \mathbf{k} \cdot \mathbf{r}_i}
    \big) \nonumber\\
    &\times \big(
    \mathfrak{U}_{\sigma_j -\mathbf k'} e^{-i \mathbf{k'} \cdot \mathbf{r}_j}
    \mathfrak{U}_{\sigma_i \mathbf k'} e^{i \mathbf{k'} \cdot \mathbf{r}_i}
    \pm \mathfrak{U}_{\sigma_i -\mathbf k'} e^{-i \mathbf{k'} \cdot \mathbf{r}_i}
    \mathfrak{U}_{\sigma_j \mathbf k'} e^{i \mathbf{k'} \cdot \mathbf{r}_j}
    \big)
    c^\dagger_{\mathbf k\uparrow} c^\dagger_{-\mathbf k\downarrow} c_{-\mathbf k'\downarrow} c_{\mathbf k'\uparrow},
\end{align}
which can be written as:
\begin{equation}
    H = \sum_{k,k'} V(\mathbf k,\mathbf k') c^\dagger_{\mathbf k\uparrow} c^\dagger_{-\mathbf k\downarrow} c_{-\mathbf k'\downarrow} c_{\mathbf k'\uparrow},
\end{equation}
and the effective BCS interaction reads as:
\begin{equation}
    V(\mathbf k,\mathbf k') = -\tfrac{U}{N} \sum_{n} \sum_{\sigma} \big(
    \mathfrak{U}^*_{\sigma \mathbf k}
    \mathfrak{U}^*_{\sigma_n -\mathbf k} e^{i \mathbf{k} \cdot \mathbf{\delta}_n}
    \pm \mathfrak{U}^*_{\sigma_n \mathbf k} e^{-i \mathbf{k} \cdot \mathbf{\delta}_n}
    \mathfrak{U}^*_{\sigma -\mathbf k}
    \big) \big(
    \mathfrak{U}_{\sigma_n -\mathbf k'} e^{-i \mathbf{k'} \cdot \mathbf{\delta}_n}
    \mathfrak{U}_{\sigma \mathbf k'}
    \pm \mathfrak{U}_{\sigma -\mathbf k'}
    \mathfrak{U}_{\sigma_n \mathbf k'} e^{i \mathbf{k'} \cdot \mathbf{\delta}_n}
    \big).
\end{equation}

On the given Fermi surface, $V(k, k')$ can be decomposed into symmetric functions $\mathfrak{V_s}(k)$, and the effective attraction $V_s$ for each symmetric channel $s$ is:
\begin{equation}
    V_s = \tfrac{1}{\mathcal{N}^2} \int_{FS} dk' \mathfrak{V_s}(k) V(k, k') \mathfrak{V_s}(k'),
\end{equation}
where:
\begin{equation}
    \mathcal{N} = \int_{FS} dk \mathfrak{V_s}(k) \mathfrak{V_s}(k).
\end{equation}
$V_s$ represents the effective attraction for different channels.

For simplicity, let us consider graphene as a simple honeycomb lattice with only nearest neighbor coupling, described by the Hamiltonian:
\begin{eqnarray}
    H_{Gr} = \begin{pmatrix}
        0 & |F(\mathbf k)| e^{i\theta(k)} \\
        |F(\mathbf k)| e^{-i\theta(k)} & 0
    \end{pmatrix},
\end{eqnarray}
where $F(\mathbf k)/t = 2 e^{-\frac{1}{2} i k_x} \cos \left(\frac{\sqrt{3} k_y}{2}\right) + e^{i k_x}$, and $|\cdot|$, $\theta(k)$ are its amplitude and phase. The energy dispersion is given by $\pm |F(\mathbf k)|$, and we consider only the positive energy branch, with wave function $\mathfrak{U}_{k} = \tfrac{1}{\sqrt{2}}(e^{i\theta(k)}, 1)$. Note that we obtain the same conclusion also for the negative energy branch. 

For the onsite attraction ($\mathcal{D} = 0$), the BCS interaction simplifies to:
\begin{equation}
    V(\mathbf k, \mathbf k') \propto -U (1 \pm 1)^2,
\end{equation}
indicating an attractive channel in the spin-singlet (plus sign) and no attraction in the spin-triplet (negative sign) channel. This leads to s-wave spin-singlet pairing instability.

\begin{figure}[t]
    \centering
    \centering\includegraphics[width=1\linewidth]{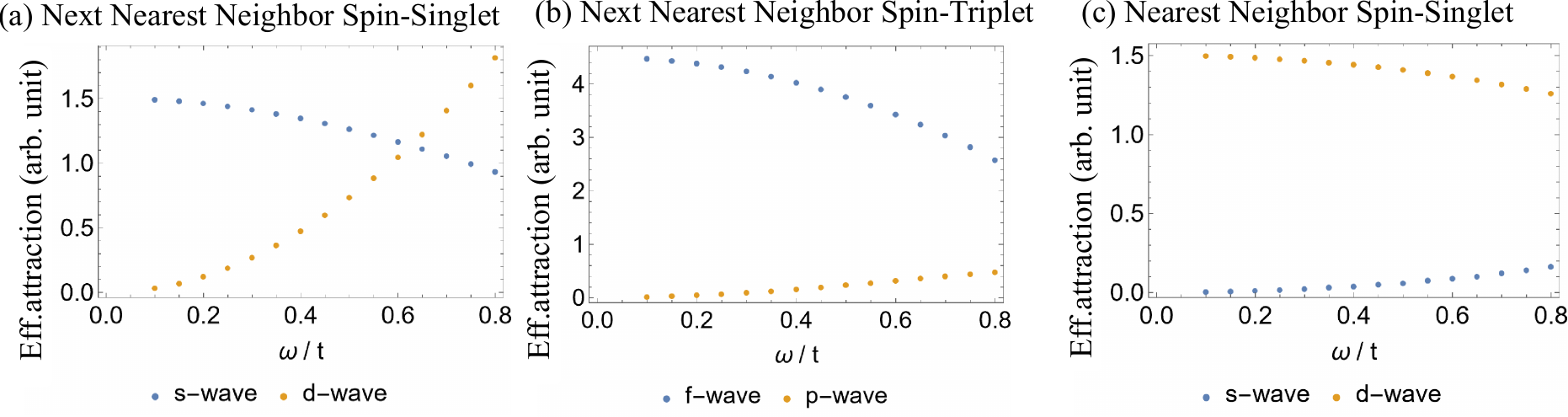}
    \caption{Effective interaction decomposition. (a)  Next nearest neighbor spin-singlet. (b) Next nearest neighbor spin-triplet. (c) Nearest neighbor spin-singlet. }
    \label{fig:SMEFFECTIVEATTRACTION}
\end{figure}

For the next-nearest neighbor spin-singlet interactions, the BCS interaction is:
\begin{equation}
    V(\mathbf k, \mathbf k') \propto -U \sum_{n} \cos (\mathbf{k} \cdot \mathbf{\delta}_n) \cos (\mathbf{k'} \cdot \mathbf{\delta}_n).
\end{equation}
We have calculated the effective attraction for s-wave and d-wave attraction channels in Fig.~\ref{fig:SMEFFECTIVEATTRACTION}(a) and found that s-wave attraction is larger near the Dirac point. In contrast, the d-wave attraction is larger near the graphene van Hove singularities. This justifies that s (d) is preferred for a higher (lower) chemical potential in Fig.3(h). 

For the next-nearest neighbor spin-triplet case, the BCS interaction is:
\begin{equation}
    V(\mathbf k, \mathbf k') \propto -U \sum_{n} \sin (\mathbf{k} \cdot \mathbf{\delta}_n) \sin (\mathbf{k'} \cdot \mathbf{\delta}_n),
\end{equation}
and as shown in Fig.~\ref{fig:SMEFFECTIVEATTRACTION}(b), the effective attraction is larger in the f-wave attraction channel. This is also consistent with the fact that in both monolayer and TBGQC with the next nearest neighbor triplet attraction, the f-wave pairing is dominant.  We also check the case of the nearest neighbor spin singlet [see Fig.~\ref{fig:SMEFFECTIVEATTRACTION}(c)] in which d-wave obtains the largest attraction. This justifies the observation that d is preferred over s in Fig.3(g). 

Therefore, the phase diagram of monolayer graphene can be understood as instability due to the largest effective attraction. Consequently, this method can give us a rough understanding of the change in the pairing symmetry in each region of the chemical potential.

When two layers are present and because the interlayer coupling is not so strong, the phase diagram resembles that of the monolayer case. The relative sign between the two superconductors can be understood by examining the Fermi surface locations. When the chemical potential is near the top or bottom of the bands, the Fermi surfaces of the two quasicrystals are centered around the $\Gamma$ point. In this scenario, it is energetically favored for the system to develop s-wave, p-wave, or d-wave pairings on both layers without requiring a change in the relative sign between the Fermi surfaces.

However, when the chemical potential is near zero energy, the Fermi pockets of the two layers are situated in different parts of the Brillouin zone, making the relative sign of pairing relevant. However, with onsite s-wave attraction, the pairing is constant across the Brillouin zone, so s-wave again remains more favorable than s*-wave.

Near the Dirac point, the two layers are effectively decoupled, meaning that the monolayer superconductors do not strongly interacting between layers, resulting in roughly degenerate pairing instability solutions between s and s*, p and p*, d and d*, and so on.

In the strong aperiodic energy regions of the QER, the situation becomes more complex as the Fermi surface is effectively disrupted, and the underlying electronic structure dictates the pairing instabilities and their relative signs. For instance, in spin-singlet channels, d-wave pairing tends to weaken in the stronger aperiodic part of QER, allowing s-wave pairing to have a higher likelihood of becoming the dominant superconducting state (see also Fig.3g [$\mu\approx -2$ eV] and Fig.~\ref{fig:figdwave}).

\section{Bott index}
\label{sec:Bott}
The Bott index serves as a topological invariant, equivalent to the first Chern number, which is already employed to investigate nontrivial states in quasicrystals. To compute the Bott index, we first obtain BdG wavefunctions.
We define the occupation projector onto wavefunctions corresponding to negative energy,
\begin{equation}\label{Eq:Projector}
	P=\sum_{\epsilon_{\lambda}<0} \left|\psi_{\lambda}\right> \left<\psi_{\lambda}\right|.
\end{equation}
In terms of this projector and $Q=I-P$, with $I$ being the identity operator, we define the projected position operators,
\begin{equation}\label{ProjectedPositionOperator}
	U_X=P e^{i 2\pi X}P+Q, \quad U_Y=P e^{i 2\pi Y}P+Q,
\end{equation}
where $X$ and $Y$ are the position operators given by
\begin{equation}\label{Eq:PositionOperator}
	X=\text{Diag}[x_1,x_1,\dots,x_N,x_N,x_1,x_1,\dots,x_N,x_N].
\end{equation}
Here, $N$ is the total number of vertices, $x_i$ is the $x$ coordinate of the $i$th vertex rescaled to $[0,1]$, and similarly for $Y$. Note that in Eq.~\ref{Eq:PositionOperator} we use the same position for both the electron and hole parts.
The Bott index is defined by
\begin{equation}\label{Eq:BottIndex}
	B=\frac{1}{2\pi} \text{Im} \big(\text{Tr}\big[\log(U_YU_XU_Y^\dagger U_X^\dagger)\big]\big),
\end{equation}
which quantifies to be a nonzero integer (zero) in topologically nontrivial (trivial) phases. As the Bott index calculation needs a large system with periodic boundary conditions (PBC), we employ graphene quasicrystal approximant with $\tau_D(g)=19/11$ which contains 5404 carbon atoms. 
We found that for energy near zero energy but finite, the same chiral of p and d always gives $|C|=2$,  while assuming opposite chiral pairing for two layers leads to a vanishing Chern number.

\clearpage
\newpage

\bibliography{refs}